\documentclass[fleqn,usenatbib]{mnras}

\let\la=\lesssim 

\usepackage[T1]{fontenc}
\usepackage{lmodern}

\DeclareRobustCommand{\VAN}[3]{#2}
\let\VANthebibliography\thebibliography
\def\thebibliography{\DeclareRobustCommand{\VAN}[3]{##3}\VANthebibliography}

\usepackage{graphicx}	
\usepackage{amssymb}	
\usepackage{amsmath}
\usepackage{subcaption}
\usepackage{newtxtext,newtxmath}
\usepackage{makecell}
\usepackage{anyfontsize}
\usepackage{pdflscape}
\usepackage{afterpage}
\usepackage{url}

\newlength{\tempdim}

\newcommand{\planck}{{\it Planck}}

\newcommand{\msun}{\hbox{$\rm ~M_{\odot}$}}

\newcommand{\kms}{km~s$^{-1}$}

\def\tdep{$t_{\rm{depl}}$}

\def\lg353{L$_{\rm 353GHz}$}
\def\ln857{L$_{\rm 857GHz}$}

\newcommand{\alphaco}{$\alpha_{\rm CO}$}
\newcommand{\alphacounits}{M$_\odot$~pc$^{-2}$~(K km/s)$^{-1}$}

\newcommand{\kb}{K$\&$B}
\renewcommand{\tabcolsep}{4pt}
 
\title[Stacking in ALCS fields]{ALMA Lensing Cluster Survey: average dust, gas, and star formation properties of cluster and field galaxies from stacking analysis}

\author[Guerrero, A. et al]{Guerrero, Andrea$^{1}$\thanks{E-mail: andreaguerrero@udec.cl},
Nagar, Neil$^{1}$,
Kohno, Kotaro$^{2,3}$, 
Fujimoto, Seiji$^{4,5}$, 
Kokorev, Vasily $^{4,5}$, 
\newauthor
Brammer, Gabriel$^{4,5}$, 
Jolly, Jean-Baptiste$^{6,7}$, 
Knudsen, Kirsten$^{6}$, 
Sun, Fengwu$^{8}$, 
Bauer, Franz E.$^{9,10,11}$, 
\newauthor
Caminha, Gabriel B. $^{12}$,  
Caputi, Karina$^{12,4}$,       
Neumann, Gerald $^{13}$,       
Orellana-Gonz\'alez, Gustavo $^{14}$,   
\newauthor
Cerulo, Pierluigi $^{15}$,          
Gonz\'alez-L\'opez, Jorge $^{16,17}$,   
Laporte, Nicolas $^{18,19}$,           
Koekemoer, Anton M. $^{20}$,             
\newauthor
Ao, Yiping, $^{21,22}$,             
Espada, Daniel, $^{23}$,            
Mu\~noz Arancibia, Alejandra M. $^{9}$
\\
$^{1}$ Departamento de Astronom\'ia, Universidad de Concepci\'on, Casilla 160-C, Concepci\'on, Chile\\
$^{2}$ Research Center for the Early Universe, University of Tokyo, Tokyo 113-0033, Japan \\
$^{3}$ Institute of Astronomy, Graduate School of Science, The University of Tokyo, 2-21-1 Osawa, Mitaka, Tokyo 181-0015,  Japan \\
$^4$ Cosmic Dawn Center (DAWN), Jagtvej 128, DK2200 Copenhagen N, Denmark \\
$^5$ Niels Bohr Institute, University of Copenhagen, Lyngbyvej 2, DK2100 Copenhagen \O, Denmark \\
$^{6}$ Department of Space, Earth and Environment, Chalmers University of Technology, Onsala Space Observatory, SE-439 92 Onsala, Sweden\\
$^{7}$ Max-Planck-Institut für extraterrestrische Physik, 85748 Garching, Germany \\
$^{8}$ Steward Observatory, University of Arizona, 933 N. Cherry Ave, Tucson, AZ 85721, USA \\
$^{9}$ Instituto de Astrof\'isica, Facultad de F\'isica, Pontificia Universidad Cat\'olica de Chile Av. Vicu\~na Mackenna 4860, 782-0436 Macul,Santiago, Chile \\
$^{10}$ Millennium Institute of Astrophysics, Nuncio Monse\~nor S\'otero Sanz 100, Providencia, Santiago, Chile\\
$^{11}$ Space Science Institute, 4750 Walnut Street, Suite 205, Boulder, Colorado 80301 \\ 
$^{12}$ Kapteyn Astronomical Institute, University of Groningen,Postbus 800, 9700 AV Groningen, The Netherlands \\
$^{13}$ Departamento de F\'isica, Universidad de Concepci\'on, Casilla 160-C, Concepci\'on, Chile \\ 
$^{14}$ Fundación Chilena de Astronomía, Santiago, Chile\\
$^{15}$ Department of Computer Science, Universidad de Concepci\'on, Concepci\'on, Chile \\
$^{16}$ N\'ucleo de Astronom\'ia de la Facultad de Ingenier\'ia y Ciencias, Universidad Diego Portales, Av. Ejército Libertador 441, Santiago, Chile \\
$^{17}$ Las Campanas Observatory, Carnegie Institution of Washington, Casilla 601, La Serena, Chile \\
$^{18}$ Kavli Institute for Cosmology, University of Cambridge, Madingley Road, Cambridge CB3 0HA, UK \\
$^{19}$ Cavendish Laboratory, University of Cambridge, 19 JJ Thomson Avenue, Cambridge CB3 0HE, UK \\
$^{20}$ Space Telescope Science Institute, 3700 San Martin Dr.,
Baltimore, MD 21218, USA \\
$^{21}$  Purple Mountain Observatory and Key Laboratory for Radio Astronomy, Chinese Academy of Sciences, Nanjing, China\\
$^{22}$ School of Astronomy and Space Science, University of Science and Technology of China, Hefei, China \\
$^{23}$ Departamento de F\'isica Te\'orica y del Cosmos, Campus de Fuentenueva, Universidad de Granada, 18071, Granada, Spain \\
}

\date{Accepted XXX. Received YYY; in original form ZZZ}

\pubyear{2021}

\begin{document}
\label{firstpage}
\pagerange{\pageref{firstpage}--\pageref{lastpage}}
\maketitle

\begin{abstract}
We develop new tools for continuum and spectral stacking of ALMA data, and apply these to the ALMA Lensing Cluster Survey (ALCS). We derive average dust masses, gas masses and star formation rates (SFR) from the stacked observed 260~GHz continuum of 3402 individually undetected star-forming galaxies, of which 1450 are cluster galaxies  and 1952 field galaxies, over three redshift and stellar mass bins (over $z = 0$--1.6 and log $M_{*} [M_{\odot}] = 8$--11.7), and derive the average molecular gas content by stacking the emission line spectra in a SFR-selected subsample. 
The average SFRs and specific SFRs of both cluster and field galaxies are lower than those expected for Main Sequence (MS) star-forming galaxies, and only galaxies with stellar mass of log $M_{*} [M_{\odot}] = 9.35$--10.6
show dust and gas fractions comparable to those in the MS.
The ALMA-traced average `highly obscured' SFRs are typically lower than the SFRs observed from optical to near-IR spectral analysis. Cluster and field galaxies show similar trends in their contents of dust and gas, even when field galaxies were brighter in the stacked maps. 
From spectral stacking we find a potential CO ($J=4\to3$) line emission (SNR $\sim4$) when stacking cluster and field galaxies with the highest SFRs.
\end{abstract}



\section{Introduction}
\label{sectintro}

Understanding how gas reservoirs and star formation rates (SFR) change with stellar mass and environment density, and evolve over cosmic time, is crucial to understand galaxy evolution. Since millimetre (mm) and submillimetre (sub-mm) observations can directly (e.g. via the CO rotational transitions) and indirectly (the sub-mm continuum) trace the gas reservoir and also the SFR
\citep[e.g.][]{Carilli2013,Scoville2013,Scoville2014,Scoville2016,Villanueva2017,Magnelli2020,Suzuki2021}, the Atacama Large Millimeter/submillimeter Array (ALMA) is a powerful tool in this area. 

The mm and sub-mm continuum trace the optically-thin Rayleigh-Jeans tail of dust emission (for typical dust temperatures of $\sim$18--50 K). 
The sub-mm flux can be used to estimate the dust mass  assuming, e.g. grey-body dust emission with two parameters: the dust emissivity index $\beta$ (in the Rayleigh-Jeans limit the grey-body flux varies as $S_{\nu} \propto \nu^{2 + \beta}$) and the dust temperature $T$. 
Values of $\beta$ are $1.5-2$ for galaxies at $z \lesssim 0.1$ \citep[e.g. ][]{Clements2010} and are consistent with $\beta= 1.8$ for galaxies at $z \sim2$--3 \citep[e.g. ][]{Chapin2009}.  
The sub-mm flux can also be used to estimate the molecular gas mass (M$_{\rm mol}$) and the mass of the interstellar medium (M$_{\rm ISM}$; atomic plus molecular gas mass; \citealt{Scoville2016}) .
Finally, the sub-mm flux can also be used to estimate the infrared luminosity (e.g. \citealt{Orellana2017}), and thus SFR, though this conversion is more uncertain than the previously mentioned conversions to dust and ISM masses, since it is more sensitive to the assumed values of the dust temperature(s) and $\beta$(s).

Star formation rates, stellar masses, and molecular gas reservoirs of galaxy populations at different redshifts help to constrain the evolution of the main sequence (MS) of star formation -- the dependence of SFR on stellar mass ($M_{*}$) -- and the star formation efficiency (SFE = SFR / M$_{\rm mol}$). At a given redshift, higher stellar mass galaxies in the MS have higher star-formation rates, and the median ratio of SFR to stellar mass (the specific SFR, or sSFR) of MS galaxies increases with redshift (e.g. \citealt{Elbaz2011}, \citealt{Rodighiero2011}, \citealt{Speagle2014}). 

Galaxies in clusters and in the field show significant differences (e.g. see reviews by \citealt{Boselli2006}, \citealt{Boselli2014}).
At low redshift, cluster galaxies tend to be more massive, older, and passive, as compared to field galaxies. 
At redshifts $z < 1$ cluster galaxies have lower mean SFRs for a fixed stellar mass, when compared to field galaxies at the same redshift  (e.g. $0.04 < z < 0.07$ \citealt{Paccagnella2016}; $0.4 < z < 0.8$ \citealt{Vulcani2010}). Low redshift cluster galaxies also have lower dust-to-stellar mass ratios than field galaxies \cite[$z\sim 0.2$]{Bianconi2020}.

Low redshift cluster galaxies typically exhibit lower molecular gas fractions than field galaxies (e.g. \citealt{Zabel2019}, \citealt{Morokuma2021}), though in some cases they can have a comparable molecular gas content \citep{Cairns2019}.
As redshift increases, there is an increase of blue, star-forming and spiral galaxies in clusters  \citep[the Butcher-Oemler effect, e.g. ][]{Butcher1978,Butcher1984,Pimbblet2003}.
At redshifts $z\gtrsim$ 1, mean SFRs increase with  environmental density in both groups (e.g. \citealt{Elbaz2007}) and clusters (e.g. \citealt{Popesso2011}).

While large-area or deep pencil beam continuum and spectral line surveys from ALMA are increasingly available (e.g. \citealt{Oteo2016}, \citealt{Walter2016}, \citealt{GonzalezLopez2017}, \citealt{Franco2018}), each has revealed relatively few individual detections of dust continuum and/or line emission. 
A promising avenue to better exploit these data is through \textit{stacking} analysis. Stacking averages the data of $N$ sources (the noise decreases by factor $\sim$$\sqrt{N}$; e.g. \citealt{Fabello2011}, \citealt{Delhaize2013}) in order to detect the average value of the stacked sources. In this regard, we could take advantage of stacking to combine the undetected signal of several galaxies in order to look for an average detection.

Stacked detections, as compared to statistics of a few individually detected sources, therefore better trace the average properties of a population. 
Finally, using stacking analysis in subsamples selected by stellar mass, redshift, and environmental density, even when individual galaxies are undetected, can provide unique constraints on the properties and evolution of the true underlying galaxy population, rather than only a few luminous or bright sources (e.g. \citealt{Coppin2015, Simpson2019,Carvajal2020}).
 
The ALMA Lensing Cluster Survey (ALCS) is the largest - in area - among the ALMA surveys targeting galaxy clusters. Combined with previous ALMA observations, it has completed observations of 33 massive galaxy clusters. All clusters have been previously imaged with the Hubble Space Telescope (HST), enabling accurate positions and other quantities derived from HST photometry. 
Initial results using the ALCS include an ALMA-Herschel study of star-forming galaxies at $z \simeq 0.5-6$ \citep{Sun2022}; the discovery of faint lensed galaxy at $z \geq 6$ \citep{Laporte2021}; a spectral stacking analysis of the undetected [C\,{\sc ii}] line in lensed galaxies at $z \sim 6$ \citep{Jolly2021}; the analysis of bright [C\,{\sc ii}] $158 \mu$m line detections of a lyman-break lensed galaxy at $z \sim 6$ \citep{Fujimoto2021}; and the discovery using the Multi Unit Spectroscopic Explorer (MUSE) of a galaxy group at $z \sim 4.3$ lensed by the cluster ACT-CL J0102-4915, also called `El Gordo' \citep{Caputi2021}.
 
In this work we apply newly developed continuum and spectral stacking tools  to ALCS maps and datacubes, in order to contrast the average dust masses, gas masses, and star formation rates in cluster versus field galaxy subsamples at $z \lesssim 1$. Given the large survey area,
relatively short integration times ($\sim$5 min) per pointing, and the large number of known optical/IR counterparts, the ALCS is one of the best existing ALMA datasets for a stacking analysis to compare the average properties of cluster and field galaxies.

This paper is organised as follows: in Section 2 we briefly describe the ALCS data and the photometric and spectroscopic catalogues used for our stacking analysis; in Section 3 we describe our methods; in Section 4 we present the results of our stacking analysis, and (sub)mm-derived dust masses, gas masses and SFRs; in Section 5 and 6 we discuss and summarise our results, respectively. 

Throughout the paper we assume a spatially flat $\Lambda$CDM cosmological model with H$_{0} = 70$ km s$^{-1}$ Mpc$^{-1}$, $\Omega_{m}$ = 0.3 and  $\Omega_{\Lambda}$ = 0.7 and the stellar initial mass function (IMF) of \citet{Chabrier2003}.

\section{Data and catalogues} \label{sec:data-cat}

We use data cubes and continuum (moment 0) maps from the ALCS (PI: K. Kohno; \citealt{Kohno2023}). The ALCS large program, with over 100 hours of integration, observed 33 massive clusters of galaxies at intermediate redshifts, between $ 0.187 < z_\mathrm{spec}  < 0.87$ (see Table \ref{table:references}). These clusters were selected from previous HST programs, including 5 galaxy clusters from the Hubble Frontier Fields (HFF; \citealt{Lotz}), 12 galaxy clusters from the Cluster Lensing and Supernova Survey with Hubble (CLASH; \citealt{Postman2012}) and 16 galaxy clusters from the Reionization Lensing Cluster Survey (RELICS; \citealt{Coe}). The ALMA survey covers a total area of 110 arcmin$^2$ (primary beam factor cut at 0.5) using a 15-GHz-wide spectral scan and reaching a typical depth of $70 \mu$Jy/beam ($1 \sigma$) at 1.2 mm. For each cluster, ALMA mosaicked a region of about $\sim$$2 \arcmin \times 2\arcmin $ around the core of the cluster. 
Spectral scans were used to cover two frequency ranges of 250.1--257.5 GHz and 265.1--272.5 GHz. For the redshifts of our clusters,  the resulting rest-frame spectra include the CO ($J=3\to2$) or CO ($J=4\to3$) line for the majority of the cluster galaxies. The wider redshift distribution of field galaxies result in rest-frame wavelengths of 250--1800 GHz, thus including more steps in the CO ladder ($J=3\to2$ and higher), and even the [C\,{\sc i}] and [C\,{\sc ii}] lines.  

In this work, for spectral stacking we use 
dirty cubes (i.e. no cleaning was performed when making the cubes) with 60 \kms\ channels and the intrinsic spatial resolution of $\sim$1--$1.5''$. 
For continuum stacking we use dirty maps (moment 0 images; created by summing all channels in the datacube) with a tapered resolution of 2\arcsec. 
All data were taken in phase-calibrated mode and using a nearby well studied phase calibrator with a highly accurate position. The positional information for any given spatial pixel in the dataset is thus accurate to $\leq0$\farcs1. 
These ALMA datasets are described in \cite{Fujimoto2023}.

Since spectral stacking requires accurate positions and redshifts to correctly align stacked emission lines (e.g \citealt{Maddox2013}, \citealt{Elson2019} , \citealt{Jolly2020}), we have compiled a single ALCS spectroscopic catalogue by combining literature spectroscopic redshifts for the 25 clusters with available data, as summarised in Table \ref{table:references}. The majority of the redshifts are from catalogues which used VLT-MUSE datasets. In constructing these catalogues, the authors used positions and sizes from imaging data to extract aperture spectra, and then used manual fitting or cross correlated with templates to identify multiple emission lines, single strong emission lines, or well defined continuum features, in order to determine a reliable or likely redshift.
From these original catalogues, we kept only meaningful extragalactic redshifts ($z_\mathrm{spec}$ $\geq$ 0), and eliminated redshifts which the authors flagged as unreliable or low quality (typically quality flag qf=1). This process resulted in a total of 9668 spectroscopic redshifts in our catalogue.

Of these a total of 2461 galaxies at redshifts $0.0 < z_\mathrm{spec} < 1.0$ and 1348 at redshifts $1.0 < z_\mathrm{spec} < 6.6$ fall inside the ALMA maps of the clusters. This subset of spectroscopic redshift sample was used in our spectral stacking analysis.
Most of the redshift compilations used here do not explicitly specify a redshift error (only a redshift quality flag). Redshifts from MUSE are expected to typically have errors of $\delta z \lesssim 0.001$ \citep{Karman2017}, i.e. a velocity error of $\lesssim 300$ \kms. 
 
\begin{table}
\caption{Spectroscopic catalogues for ALCS Clusters} 
\centering 
\begin{tabular}{cccc}
\hline 
HST field  &Cluster             &   $z_\mathrm{spec}$     &    References    \\ [0.5ex]
\hline                                                      
    & Abell 2744        & 0.308       &   \cite{Richard2021}   \\
    & Abell S1063       & 0.348       &   \cite{Karman2017}    \\ 
    & & &  \cite{Mercurio2021}  \\
HFF    & Abell370          & 0.375    &   \cite{Richard2021}   \\
       & MACS J0416.10-2403& 0.396       & \cite{Richard2021}  \\ 
        & MACS J1149.5+2223 & 0.543      & \cite{Grillo2016}   \\
                                     & & & \cite{Treu2016}  \\\hline
        & Abell383           & 0.187     & \cite{Geller2014}     \\ 
         & Abell209          & 0.206     & \cite{Annunziatella2016} \\ 
        & RXJ2129.7+0005     & 0.234     & \cite{Jauzac2020}     \\ 
         & MACS1931.8-2635   & 0.352     & \cite{Caminha}        \\ 
         & MACS1115.9+0129   & 0.352     & \cite{Caminha}        \\
         & MACS0429.6-0253   & 0.399     & \cite{Caminha}        \\
CLASH    & MACS1206.2-0847   & 0.440     & \cite{Richard2021}    \\
         & MACS0329.7-0211   & 0.450     & \cite{Richard2021}    \\ 
         & RXJ 1347-1145     & 0.451     & \cite{Richard2021}   \\      
         & MACS1311.0-0310   & 0.494     & \cite{Caminha}      \\ 
         & MACS1423.8+2404   & 0.545     & \cite{Treu2015}      \\ 
                                     & & & \cite{Schmidt2014}    \\
         & MACS2129.4-0741   & 0.570     & \cite{Jauzac2020} \\ \hline 
         & Abell 2163         & 0.2030   & \cite{Rescigno2020}  \\ 
         & PLCK G171.9-40.7   & 0.2700   &       -        \\ 
        & Abell 2537        & 0.2966     &   \cite{Foex2017}  \\ 
        & AbellS295         & 0.3000     &  \cite{Bayliss2016} \\ 
         & MACSJ0035.4-2015  & 0.3520    &      -        \\ 
        & RXC J0949.8+1707  & 0.3826     &      -        \\ 
        & SMACSJ0723.3-7327  & 0.3900    &      -       \\ 
RELICS & RXC J0032.1+1808  & 0.3956      &      -    \\   
        & RXC J2211.7-0350   & 0.3970    &      -             \\  
       & MACSJ0159.8-0849  & 0.4050      &    \cite{Stern2010}                  \\  
       & Abell 3192        & 0.4250      &      -               \\  
      & MACSJ0553.4-3342   & 0.4300      &  \cite{Ebeling2017}  \\
     & MACSJ0417.5-1154   & 0.4430       &  \cite{Jauzac2019}   \\ 
      & RXC J0600.1-2007   & 0.4600      &      -             \\
       & MACSJ0257.1-2325   & 0.5049     &  \cite{Stern2010}  \\  
      & ACT-CLJ0102-49151  & 0.8700      &  \cite{Sifon2013} \\ 
\hline 
\end{tabular}
\label{table:references} 
\end{table}

\begin{figure*}
    \centering
    \includegraphics[width = 0.8\textwidth]{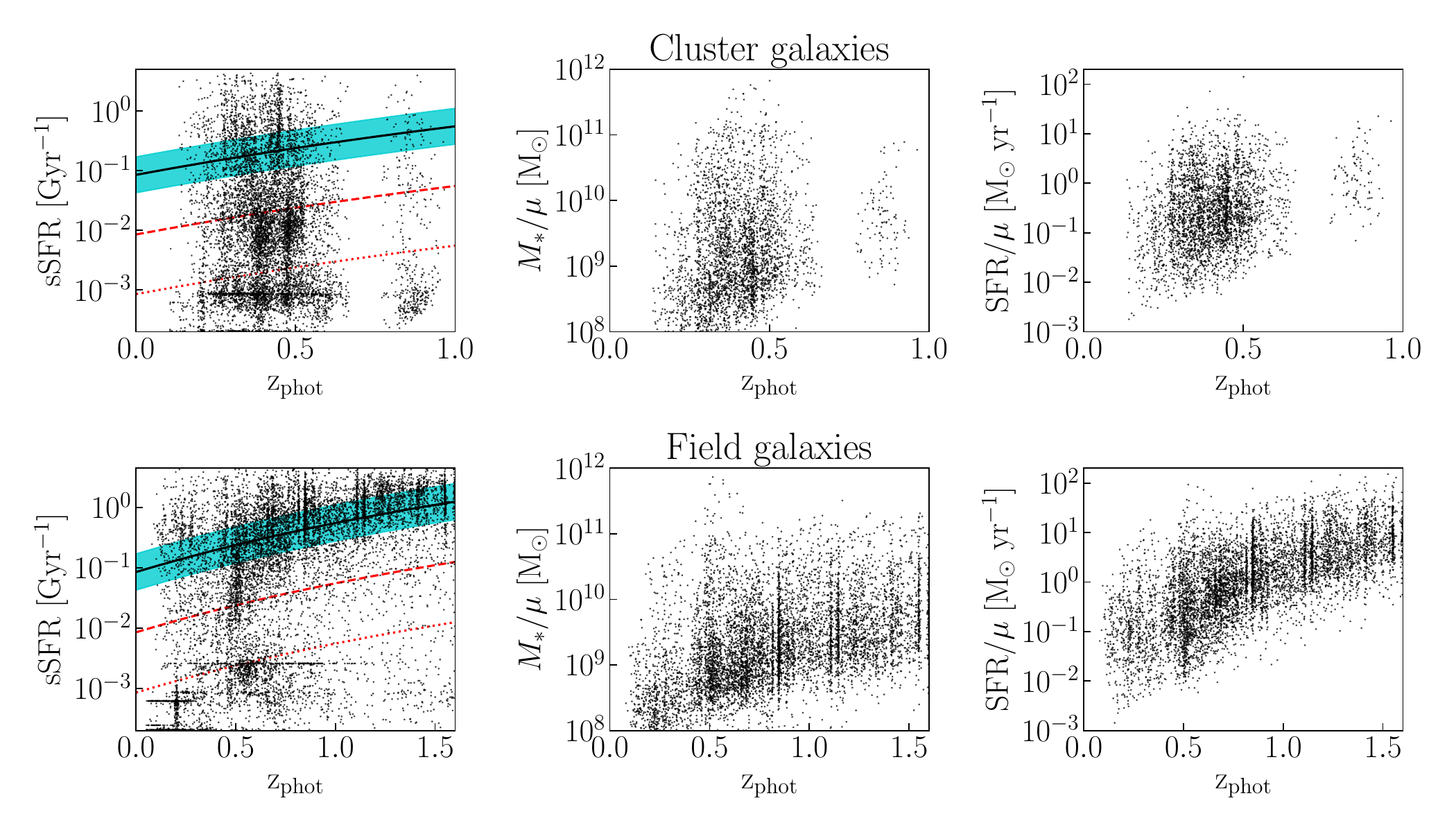}
    \caption{Left panels: Specific SFR (sSFR) as a function of photometric redshift for all cluster (top) and field (bottom) galaxies in our parent sample with all filters applied except for filter (viii) described in Sect.~\ref{sec:data-cat}. 
    The black line and cyan shaded region show the predicted redshift evolution of the main sequence (MS) of star formation from \citet{Elbaz2011}, and its $\sim$0.3 dex dispersion. The red dashed lines trace 1 dex below the MS; galaxies above this line are in our  star forming galaxies (`ALCS-SF') sample used in the stacking analysis. The dotted red lines trace 2 dex below the MS; galaxies below this line are not used in our analysis. Galaxies between the two red lines are considered passive galaxies.
    Middle and right panels:  stellar masses (middle) and SFRs (right panels), as a function of photometric redshift, for the `ALCS-SF' sample, i.e. galaxies above the dashed red line in the left panels. }
    \label{fig:sm-sfr-ssfr}
\end{figure*}

For continuum stacking of the ALCS ALMA datasets we use the version 1.0 (v1.0) photometric catalogues of \citet[][hereafter \kb\  catalogues]{kokorev22} which include all 33 ALCS clusters. 
The v1.0 \kb\ catalogues apply the EAZY code \citep{Brammer2008} to HST and Spitzer photometry to derive photometric redshifts. The best-fitting templates to the observed-frame UV-to-near-IR photometry are also used to derive stellar masses and SFRs. The full photometric catalogue includes $\sim$200,000 sources at redshifts $0 < z_\mathrm{phot} < 12$. The \kb\ catalogue photometry was typically extracted over 0\farcs7 to $\sim$3\farcs5 diameter apertures, roughly comparable to the 1\arcsec\ resolution ALMA continuum images used by us and smaller than the 20\arcsec\ stamp-size we use in our continuum stacking.

Using the \kb\ catalogue for continuum stacking, instead of the spectroscopic redshift catalogue, gives us two advantages: (a) a larger number of targets to stack and, therefore, a higher signal to noise ratio in the stacked images. Note that the photometric redshift errors are typically smaller than the redshift bins we use for stacking; (b) the stellar masses and SFRs derived in these catalogues allow us to stack in bins of stellar mass and SFR. Note that these quantities are derived from template fitting, and it is nontrivial to scale them to a different (spectroscopically derived) redshift.

The \kb\ catalogues provide magnification factors for lensed sources; their listed fluxes, SFRs, and stellar masses are not corrected for this magnification. We use these magnification factors to derive the demagnified stellar mass, SFR and flux for each object. Therefore, henceforth all physical properties from the \kb\ catalogues refer to the demagnified quantities. 

 We refer the reader to \citet{kokorev22} for a full comparison between their photometric redshifts and the $\sim$7000 spectroscopic redshifts available in their fields. They found a good agreement of the two in $\sim$80\% of their sample. The remaining $\sim$20\% have large, sometimes catastrophic $\mid$ $z_{\rm phot}$ - $z_{\rm spec}$ $\mid$ $\sim3$, errors in photometric redshifts. These photometric redshift failures are mainly due to confusion of the Lyman, Balmer, and 4000\AA\ breaks in the template SEDs. The catastrophic redshift errors result in high redshift galaxies being assigned a photometric redshift $z_{\rm phot} \sim0$, and vice-versa.

Using the full \kb\ catalogue for our cluster and field galaxies produces noticeable `striping' when plotting stellar mass as a function of redshift, likely due to erroneous photometric redshifts. 
To avoid galaxies with `catastrophic redshift errors' from the \kb\ catalogue (which would produce erroneous stellar masses and SFRs), to more reliably separate cluster and field galaxies, to avoid strong line-emitting galaxies producing false continuum detections, and to eliminate passive galaxies, we apply the following filters to the catalogue:
\begin{enumerate}
\item  Magnitude cutoff of 24 in the H-band, to minimise contamination caused by blue faint galaxies.
\item Discarding galaxies without IRAC photometry available (flagged as \texttt{bad$\_$phot} = 1), to avoid selecting galaxies with low quality photometry.
\item  Selection of galaxies with |$z_\mathrm{err}| / z_\mathrm{phot} < 0.5$, where $z_\mathrm{err}$ is the `16 percentile to 84 percentile error' of $z_\mathrm{phot}$ listed in the catalogues. Here we do not use the typical |$z_\mathrm{err}$| / (1+$z_\mathrm{phot}$) criterion since we later require to reliably separate cluster and field galaxies.
\item Selection of galaxies with $z_\mathrm{phot} >0.05$, to avoid $z \sim 4$ galaxies for which $z_\mathrm{phot} \sim 0$ (see above).
\item Using a lensing magnification cutoff of $\mu < 20$, to avoid galaxies that may have an erroneous large magnification factor.
\item Discarding galaxies with individual emission line detections (\citealt{Fujimoto2023}, Gonzalez et al., in prep.). 
\item Selection of galaxies with $M_*$ > $1\times 10^{8}$ $M_{\odot}$ to eliminate galaxies with erroneously low stellar masses due to erroneous redshifts.
\item Selection of galaxies with sSFR $\geq$ (MS -- 1 dex), i.e. higher than 1 dex below the expected MS, to eliminate passive galaxies and to eliminate galaxies with erroneously low SFR due to erroneous redshifts. The typical 1$\sigma$ spread of the Main Sequence is $\pm$0.3 dex (e.g. \citealt{Elbaz2011}). Given the likely larger errors in \kb\ catalogue SFR and stellar masses (due to these depending on  photo-z determinations), and the clear clump of passive galaxies seen in the top left panel of Fig. \ref{fig:sm-sfr-ssfr}, we use a larger spread of 1 dex (above dashed red line) in order to capture most star-forming galaxies below the fiducial MS. As seen in the left panels of Fig. \ref{fig:sm-sfr-ssfr}, this sSFR selection eliminates most passive galaxies in the cluster galaxy sample.
\end{enumerate}

After applying these filters, the catalogue has $\sim$13000 sources, at redshifts $0 < z_\mathrm{phot} < 12$.

We then divided the filtered photometric catalogue into cluster and field galaxy sub-catalogues at $z_\mathrm{phot} \leq1.6$. 
Cluster galaxies were selected as those at $z_\mathrm{phot} \pm 0.1$ from the cluster redshift, and field galaxies as those that do not fulfil this condition. The $\pm 0.1$ range is used to avoid significant contamination of the galaxy cluster subsample given that the photometric redshift errors of the filtered sample at $z_\mathrm{phot} \leq1.6$ is |$z_\mathrm{err}| / z_\mathrm{phot} < 0.1$.  

The final catalogue we use is thus left with 3321 cluster galaxies at $0.0<z_\mathrm{phot} \leq1.0$ and 7421 field galaxies at $0.0<z_\mathrm{phot} \leq1.6$. We define this filtered catalogue of star-forming galaxies as `ALCS-SF'. The left panels of Fig. \ref{fig:sm-sfr-ssfr} show the distributions of sSFR for all galaxies in our sample (but not yet applying filter (viii) above), to clearly distinguish the star-forming galaxies we will use in the stacking analysis (above the dashed red line), the passive galaxy sample (between the two red lines), and the galaxies we do not use (below the dotted red line). The middle and right panels show the stellar mass and SFR as a function of photometric redshift, for the cluster (top) and field (bottom) galaxies selected after all filters (including filter (viii)) were applied, i.e. the `ALCS-SF' sample.

In the `ALCS-SF' sample, $1656$ cluster galaxies and $2223$ field galaxies fall inside the ALMA maps of the clusters. These ALMA-observed galaxies were further divided into stellar mass and redshift bins. 

Details on this binning, and the final number of galaxies stacked, can be found in Section \ref{sect:flux_dust_ISM}. A comparison of spectroscopic and photometric redshifts of the galaxies used in the stack gives a median error of |$z_\mathrm{spec} - z_\mathrm{phot}$ | / $z_\mathrm{spec}$ = 0.07, justifying our use of the $z \pm 0.1$ criterion for cluster galaxies.

Our continuum-stacking results primarily contrast cluster and field galaxies at $0.0<z_\mathrm{phot} \leq1$; field galaxies at $1<z_\mathrm{phot} \leq1.6$ are only used to calibrate the redshift dependence of the scaling relations we use for converting stacked radio fluxes into physical quantities. A stacking analysis of sources at $z_\mathrm{phot} \geq1$ will be presented by Jolly et al. (in prep). 

\section{Methods}
\subsection{Stacking software}

We have developed continuum and spectral stacking codes, which are made public here. Both have to be executed within the Common Astronomy Software Applications package (CASA; \citealt{McMullin}). Both primarily rely on python packages included in CASA and are thus easy to execute in current and future versions of CASA.

Publicly available stacking softwares for ALMA data include the continuum (in image and \textit{uv}-domain) stacking package of \cite{Lindroos2015} and the spectral stacking package of \cite{Jolly2020}. Our new stacking codes are based on the \cite{Lindroos2015} code - specifically, we use their functions to handle the input maps/cubes and the stacking positions - but with several modifications in order to optimise for the case of incremental stacking of very large datasets, where stacking requires to be rerun whenever a single or a few new maps/cubes are added to the overall large set of input maps/cubes. This is handled by storing intermediate results in the stacking process; when the stacking is rerun, the user has the option to use these intermediate results together with the few new apertures or images to be extracted before the final stack. The stored intermediate results are useful for posterior analysis. For example, in the case of image stacking, all extracted sub-images are stored in an ALMA cube format. This allows the user to easily stack subparts of the `cube' or use Monte Carlo (MC) analysis to test the robustness of the stacked result in case of e.g. position errors or to test the effect of a single to a few subimages on the stacked result. 

At the end of the stacking process, plotting scripts provide easy to visualise plots and analyses of the intermediate and final extracted images or spectra.
The stacking software is detailed in Appendix \ref{sectappendixstack} and its documentation is available on GitHub\footnote{\url{https://github.com/guerrero-andrea/stacking_codes}}.

\subsection{Star Formation Rates, Dust masses and ISM masses from ALMA 1.2~mm fluxes} \label{met:sfr-dust-gas}

\begin{figure}
    \centering
    \includegraphics[width = 1\columnwidth]{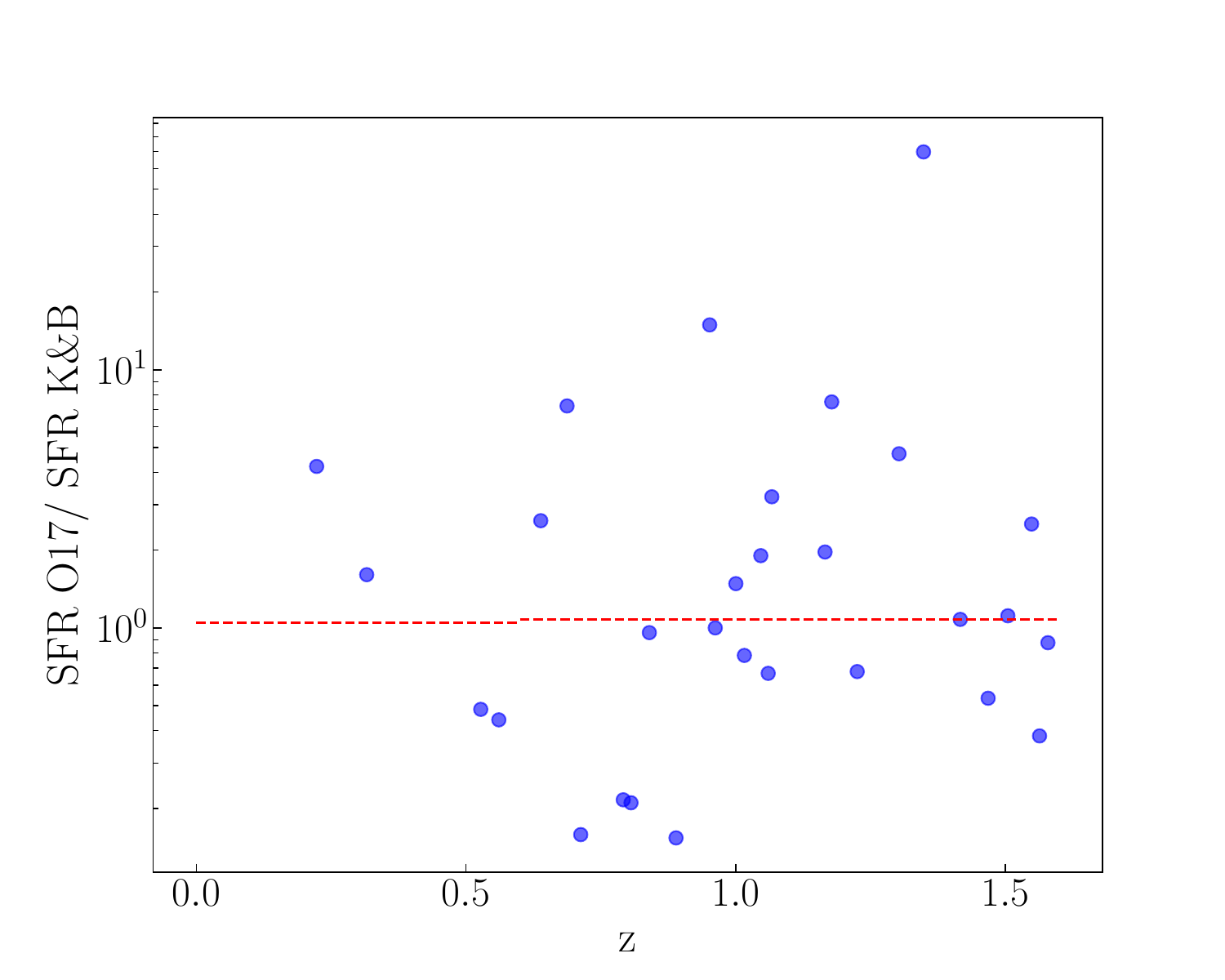}
    \caption{
    Ratio between SFRs derived from a single ALMA flux using our method (based on equations \ref{eq:greybody} to \ref{eq:sfr}) and the SFRs listed in the \kb\ photometric catalogues, for individually detected sources from \protect\cite{Fujimoto2023}.
The sample was divided into two bins and we find the best median agreement when using T $= 22.8$ K for sources at  $0.0 < z \leq 0.6$ and T $ = 22.4$ K for sources at $ 0.6 < z \leq 1.6$. The median value for each bin is shown with the red dashed lines.}
    \label{fig:sfr-o17-cats}
\end{figure}

We derive SFRs from the observed-frame 1.2~mm ALMA fluxes (individual fluxes of detected sources or fluxes in stacked maps). Fluxes in individual or stacked maps are measured using the source extraction software \textsc{Blobcat} (see Section \ref{sect:flux_dust_ISM}). The measured flux is immediately corrected for lensing magnification using the magnification listed in the \kb\ catalogue for an individual source, or the mean magnification of all sources in the stack for stacked maps.

First, we use the photometric redshift (individually detected sources) or mean photometric redshift of the stacked sample (stacked detections) to derive a representative rest wavelength [1.2mm/(1+$<$z$>$)] for the map and test if this is closer to $350\mu$m or $850\mu$m, the two wavelengths at which we have a reliable flux to SFR conversion from \cite{Orellana2017}. The measured flux (the total flux from a 2D Gaussian fit, see Sect.~\ref{sect:flux_dust_ISM})
 in the map is then extrapolated to the closest of the above two wavelengths assuming a grey body spectrum;
\begin{eqnarray}
    I_{\nu} = \textup{B}_{\nu}(T_{\rm d}) \nu^{\beta}
\label{eq:greybody}
\end{eqnarray}
where we choose $\beta = 1.8$; this value for $\beta$ is supported in several samples of low and high redshift galaxies (e.g. \citealt{Scoville2016,Scoville2017,Dunne2022}). Using values of $\beta = 1.2$ instead would change our flux extrapolations by $\sim$5--30\% over the redshift range we use. The luminosity-weighted dust temperature (T$_{\rm d}$) is calibrated further in this subsection.

The extrapolated flux at $350\mu$m or $850\mu$m is then converted into a specific luminosity $L_{\nu}$ [W Hz$^{-1}$] at either $\nu = 857$ GHz or $353$ GHz. This specific luminosity is, in turn, converted to an infrared luminosity $L_{\textup{IR}}$, and then to SFR, using the equations of \citet{Orellana2017}:

\begin{eqnarray}
\log \left( \frac{L_{\textup{IR}}}{[L_{\odot}]} \right) &=& 1.017  \log \left( \frac{L_{350\mu m}}{[\textup{W Hz}^{-1}]} \right) \nonumber \\ 
&+& 0.118  \left( \frac{T_{\textup{cold, dust}}}{[K]} \right) - 16.45 \label{eq:Lir1}\\ 
\log \left( \frac{L_{\textup{IR}}}{[L_{\odot}]} \right) &=& 1.01  \log \left( \frac{L_{850\mu m}}{[\textup{W Hz}^{-1}]} \right) \nonumber \\ 
&+& 0.15  \left( \frac{T_{\textup{cold, dust}}}{[K]} \right) - 15.93  \label{eq:Lir2}
\end{eqnarray}
\begin{eqnarray}
\mathrm{SFR} [M_{\odot} \textup{yr}^{-1}] = 1.05 \times 10^{-10} L_{\textup{IR}} [L_{\odot}].
\label{eq:sfr}
\end{eqnarray}

These \citet{Orellana2017} calibrations were based on a large sample of redshift zero galaxies with \planck\ flux measurements. Our use of 1.2~mm observed frame fluxes, thus rest-frame fluxes at $\geq$600\micron, justifies the use of T$_{\rm d}$ = T$_{\rm cold,dust}$ for our galaxies or stacks, 
though we caution the reader these zero redshift calibrations are being used here to derive SFRs out to redshift 1.6. A justification of the value of T$_{\rm d}$ used in these equations is provided further below in this section.
We further
note that \citet{Orellana2017} used the relationship of \citealt{Kennicutt1998} and a \cite{Salpeter1955} IMF. We divided their equation by factor 1.7 (e.g. \citealt{Genzel2010}, \citealt{Man2016}, \citealt{Lagana2016}) in order to base our SFR (eqn. \ref{eq:sfr}) on a \cite{Chabrier2003} IMF. 

\begin{figure*}
\centering
\begin{minipage}[b]{0.3\textwidth}
\begin{subfigure}[b]{\linewidth}
\centering
\includegraphics[width=0.7\linewidth]{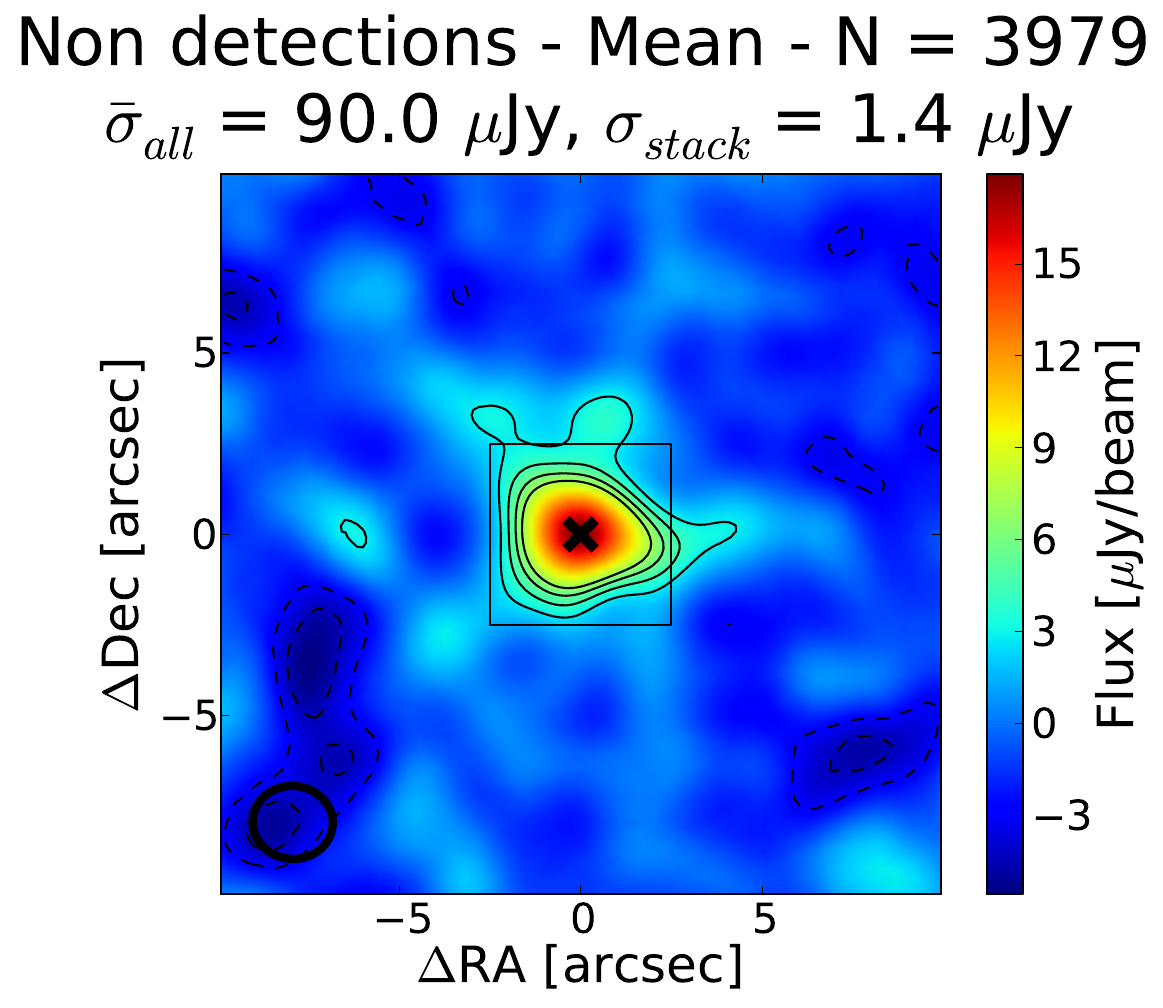}
\end{subfigure}\\[\baselineskip]
\begin{subfigure}[b]{\linewidth}
\centering
\includegraphics[width=0.7\linewidth]{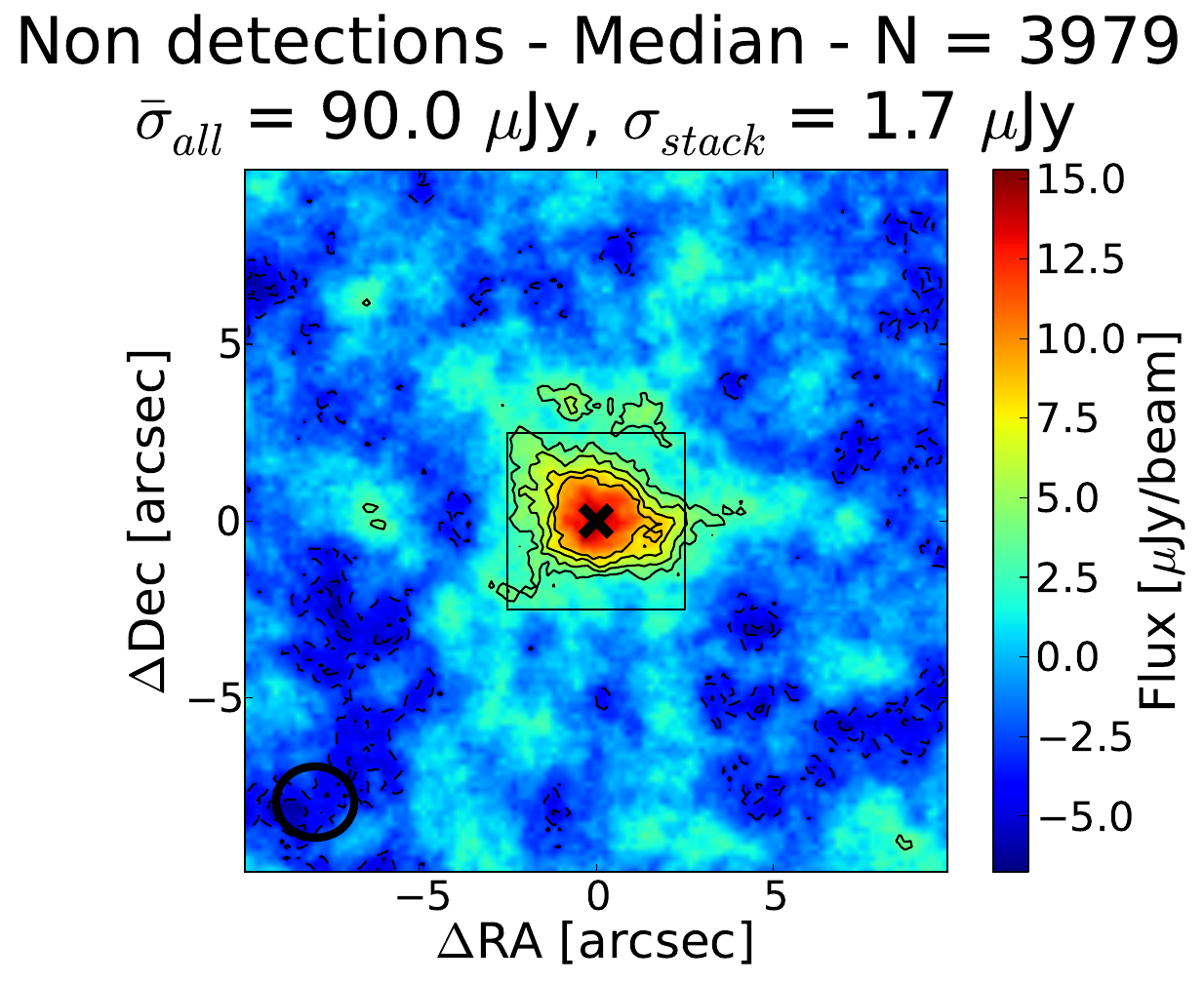}
\end{subfigure}
\end{minipage}
\hfill
\begin{subfigure}[b]{0.65\textwidth}
\centering
\includegraphics[width=0.9\linewidth]{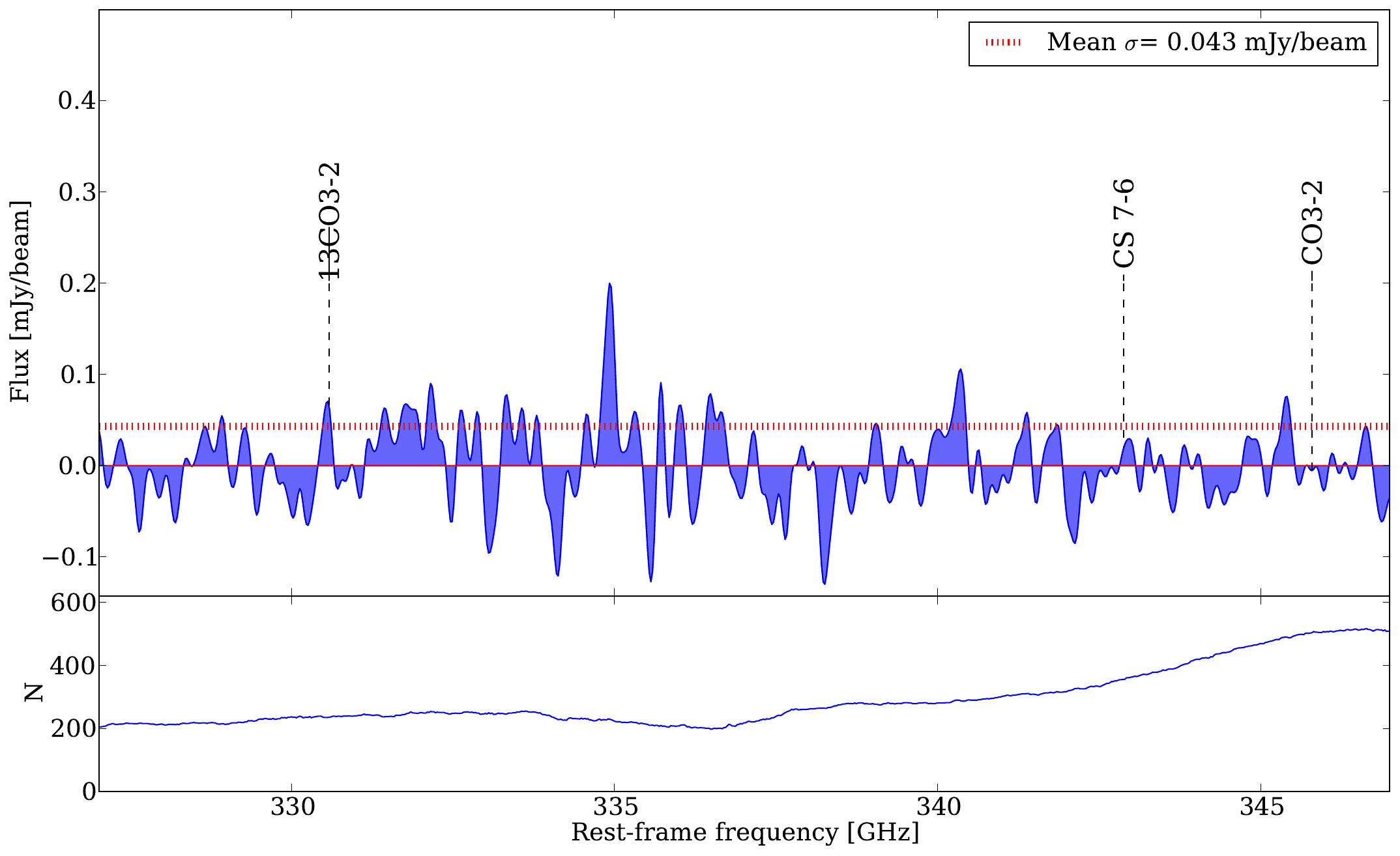}
\end{subfigure}
\caption{Left panels: continuum stacks of all undetected sources from the \kb\ catalogue. Right panel: spectral stack of all undetected sources from our compiled spectroscopic catalogue. The left upper (lower) panel shows the mean (median) continuum stack of the individually undetected sources. Above each panel, we show the number of sources stacked (N), the average rms of all input maps that were stacked ($\bar{\sigma}_{all}$) and the rms of the final stacked map ($\sigma_{stack}$). The black crosses show the centre of each map and the black circle shows the synthesised ALMA beam size. The  black contours show the levels of [$-3\sigma,-2\sigma,2\sigma,3\sigma,4\sigma$], where $\sigma$ is the standard deviation of the edge of the image (outside the black square). The right panel shows an excerpt of the frequency range from a mean spectral stack of all sources with redshifts in our compiled spectroscopic catalogue, which have no individually detected emission lines. The upper panel shows the stacked spectrum as a function of the rest-frame frequency. The dashed red line shows the mean standard deviation $\sigma$ of the spectrum. The lower panel shows the number of objects stacked at each frequency.}
\label{fig:all-spectral-continuum-stacking}
\end{figure*}

Since we will use these equations in stacked fluxes, we tested if the estimated SFR from stacked galaxies (i.e. using the mean flux and mean redshift to derive a mean/stacked SFR) is equivalent to computing the SFR per galaxy and then taking the mean (i.e. the true mean SFR). A full description of this test can be found in Appendix \ref{appendix:sfr}.
For this test, we use MC simulations to compare between the derived SFR for a source with constant flux as a function of redshift, the derived mean SFR from a mean flux and redshift of random sources (stacked SFR), and the derived mean SFR from individual SFR measurements per random source (true mean SFR). For the last two cases, each source is simulated to have a random flux between 0 and 0.1 mJy, which are typical values of noise within the continuum images of 1.2mm. These scenarios are shown in Fig. \ref{fig:sfr-z-no550} (top panel). 
These MC results can be compared to those for a source with a constant flux of 0.05 mJy (the mean value of the random fluxes used) over the full redshift range. As an additional comparison, we show how the true mean SFR of individually detected galaxies from \cite{Fujimoto2023} was compared to the SFR value derived from their continuum stacked image.

In Fig. \ref{fig:sfr-z-no550} we used a constant dust temperature with redshift. For this test, we used T$_{\rm d}=22$K, which is within the temperature range seen in both local star forming \citep{Dunne2022} and in quiescent galaxies \citep{maget21}, although any temperature in this range could have been used. Clearly, for $z \gtrsim 1$, the estimated SFR (for the constant flux) is relatively independent of redshift: thus the true mean SFR and the stacked SFR are relatively commutative at these redshifts. However, at $z \lesssim 1$, the SFR of a constant flux source varies significantly with redshift, so the mean SFR of several detected sources is not necessarily equivalent to the SFR obtained by first stacking the maps, and then using the average flux and redshift to obtain the stacked SFR. Nevertheless, we argue that this SFR derivation can also be used at z $\lesssim$ 1 for the following reasons.

The redshift distributions of each stellar mass sub-bin of each redshift bin are not significantly different in most bins (each bin description is found in Section \ref{sect:flux_dust_ISM}). In the absence of strong systematic effects (evolution of temperature or SFR with redshift) we would expect that the process of stacking and then estimating the SFR via the mean redshift is not too different from first estimating individual SFRs and then later averaging. In fact, Fig. \ref{fig:sfr-z-no550} shows that, for these redshift distributions, the mean SFR of the galaxies in the sample is fairly similar to the stack-derived mean SFR and remains within $3\%$ for the simulated sources and $30 \%$ for detected galaxies, under the assumption of no systematic effects. 

\begin{figure*}
    \centering
    \begin{subfigure}[t]{0.475\textwidth}
        \centering
        \includegraphics[width = 0.98\columnwidth]{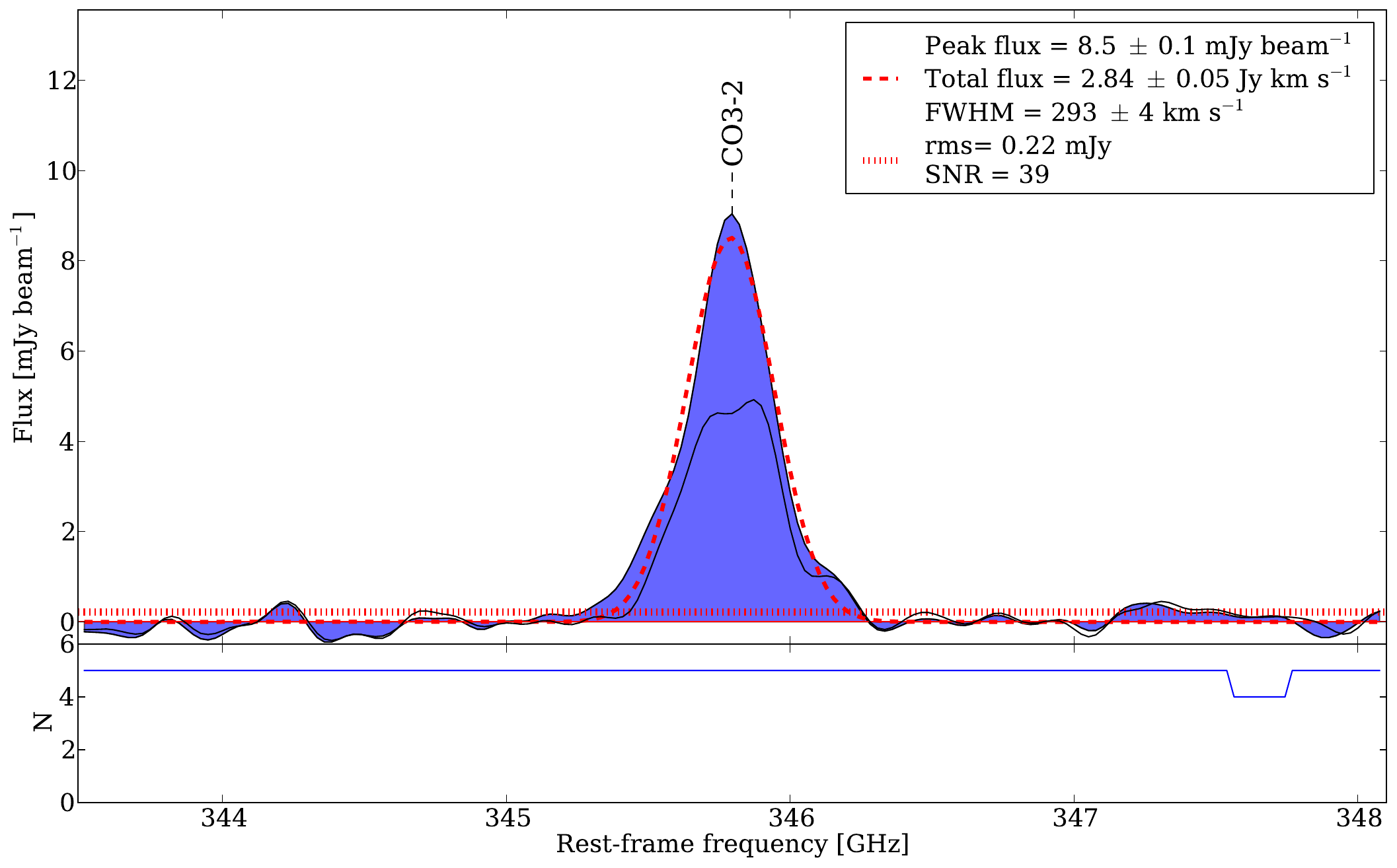}
    \end{subfigure}
    \begin{subfigure}[t]{0.475\textwidth}
        \centering
        \includegraphics[width = \columnwidth]{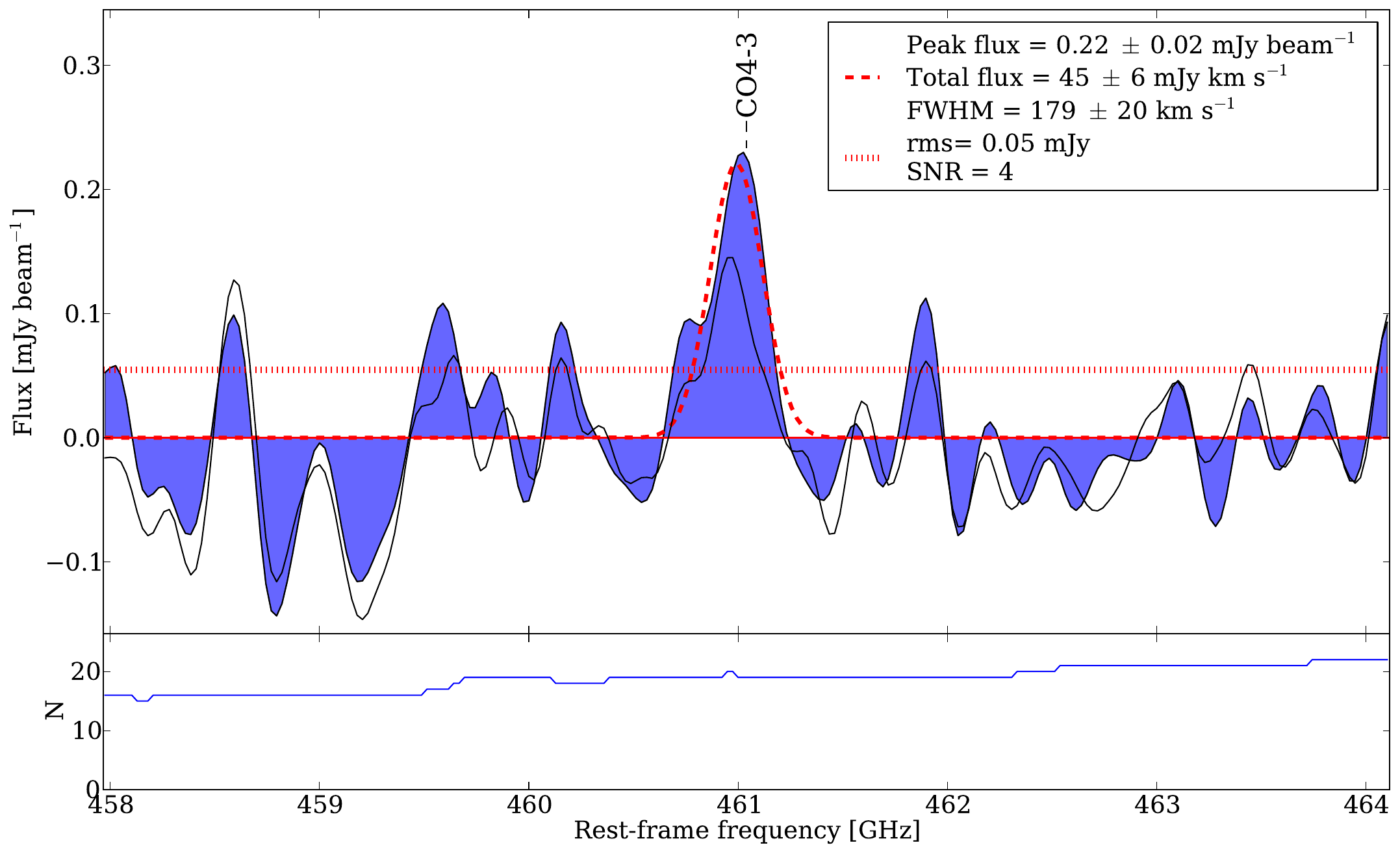}
    \end{subfigure}%
\caption{
Spectral stacking results: 
The left panel shows the stacked CO ($J=3\to2$) spectrum of all (5) sources for which redshifts are available in our compiled spectroscopic catalogue and in which this line is individually detected. The right panel shows the spectrum of the CO ($J=4\to3$) line in a spectral stack of sources in our compiled spectroscopic catalogue which 
individually show no detected emission lines, and are in the uppermost 15-percentile SFR bin, where the SFRs were taken from the \kb\ catalogue.
In both panels, the mean stacked profiles are shown in blue, the best-fitting Gaussian to this profile in red, and the median stack line profile in black. The bottom sub-panels show the number of objects stacked.
}
    \label{fig:spectral-stack-subsamples}
\end{figure*}

The remaining unknown is thus the (single dust component) temperature, T$_{\rm d}$ to be used in equations \ref{eq:greybody}, \ref{eq:Lir1} and \ref{eq:Lir2}. 
To calibrate this temperature (and its variation with redshift) we compared our derived SFRs to those listed in the \kb\ catalogue, under the assumption that the unobscured and obscured star formation are correlated.
For this we selected the continuum sources individually detected in the ALCS fields at signal to noise ratios (SNR) $> 5$ \citep{Fujimoto2023}, and classified as `ALCS-SF' galaxies in our sample, and used these fluxes to derive the `highly obscured' SFR via our method (eqns. \ref{eq:greybody} to \ref{eq:sfr}; hereafter the `O17' method). We first searched for the galaxies that were also in the \kb\ catalogue, and we found 53 sources. Then we selected only the ones with redshift $z\leq1.6$, resulting in 29 galaxies.
We divided the sample into two redshift bins, $0<z\leq0.6$ and $0.6<z\leq1.6$. For each bin, we assume a single temperature and derive the O17 SFR for the galaxies. 
We varied T$_{\rm d}$ in order to get a median ratio close to 1 for the O17 to \kb\ SFRs. 
This resulted in  T$_{\rm d} = 22.8$ K for the lowest redshift bin, and T$_{\rm d} = 22.4$ K  for the second redshift bin. 
Fig. \ref{fig:sfr-o17-cats} shows the ratio of our O17-derived SFRs (derived from a single observed 1.2~mm flux) to the \kb\ catalogue SFRs. 

A detailed analysis of SF galaxies and SMGs \citep{Dunne2022} found SF galaxies to have mass-weighted temperatures (T$_{\rm mw}$) in the range  20--25 K with a median of 23 K, a value slightly lower than the 25 K used by \citep{Scoville2016} in a mixed sample of SFs and SMGs. The luminosity weighted dust temperature (T$_{\rm d}$) is expected to be higher than the mass-weighted equivalent (T$_{\rm mw}$) by a few degrees \citep[e.g.][]{Scoville2016}, but in the absence of detailed multi-component dust temperature fits, we assume that T$_{\rm d}$ = T$_{\rm mw}$.  
Hereafter we use a value of 23 K for all of T$_{\rm d}$,   T$_{\rm cold,dust}$, and T$_{\rm mw}$, and note that this temperature is slightly higher than the luminosity weighted temperatures ($\sim$19--21 K) seen in quiescent galaxies over this redshift range \citep[e.g. Fig. 6 of][]{maget21}.  
Varying T$_{\rm d}$ by $\pm$2 K results in +100\% and $-$50\% changes in the derived SFRs via eqns. \ref{eq:greybody} to \ref{eq:sfr}.

We estimate dust masses using the following equations from \citet{Orellana2017} with T$_{\rm d}$ = 23 K: 

\begin{eqnarray}
\log \left( \frac{M_{\textup{dust}}}{[M_{\odot}]} \right) &=& 0.940  \log \left( \frac{L_{350\mu m}}{[\textup{W Hz}^{-1}]} \right) \nonumber \\ 
&-& 0.0791  \left( \frac{T_{\textup{cold, dust}}}{[K]} \right) - 12.60 \label{eq:Md1}\\ 
\log \left( \frac{M_{\textup{dust}}}{[M_{\odot}]} \right) &=& 0.993  \log \left( \frac{L_{850\mu m}}{[\textup{W Hz}^{-1}]} \right) \nonumber \\ 
&-& 0.054  \left( \frac{T_{\textup{cold, dust}}}{[K]} \right) - 13.310  \label{eq:Md2}
\end{eqnarray}
Varying T$_{\rm d}$ by $\pm$2 K would result in a $\pm$ 30\% change in the dust mass in the inverse sense. 

We derived dust-corrected near ultraviolet (NUV) SFRs using NUV fluxes from the \kb\ catalogues and IR luminosities derived from the 1.2mm ALMA fluxes (eqns \ref{eq:Lir1} and \ref{eq:Lir2}). For this we followed the method described by \cite{Hao2011}. The corrected NUV luminosity, L(NUV)$_{\mathrm{corr}}$, is estimated as,
\begin{eqnarray}
\mathrm{L(NUV)}_{\mathrm{corr}} = \mathrm{L(NUV)}_{\mathrm{obs}} \mathrm{[erg \ s^{-1}]}+ 0.27 L_{\textup{IR}} \mathrm{[erg \ s^{-1}]}
\end{eqnarray}

Where $\mathrm{L(NUV)}_{\mathrm{obs}} \mathrm{[erg \ s^{-1}]}$ is measured from the NUV flux densities using $L = \nu\,L_\nu$.
Then, the corrected SFR is measured as,
\begin{eqnarray}
\log \mathrm{SFR}_{\mathrm{corr}} [M_{\odot} \textup{yr}^{-1}]  = \log  \mathrm{L(NUV)}_{\mathrm{corr}} - 42.959
\end{eqnarray}

Finally, the value of SFR$_{\rm corr}$ derived above is divided by a factor 1.7 in order to convert a Salpeter based SFR into a \cite{Chabrier2003} based SFR.

This single-component dust temperature of 23 K is also used, together with the observed 1.2 mm flux, to estimate the molecular gas mass ($M_{\textup{mol}}$) using eqns. A6, A14 and A16 of \citet{Scoville2016,Scoville2016b}, and using $\alpha_{\rm 850}$ = 5.5 $\times10^{19}$ erg s$^{-1}$ Hz$^{-1}$ $M_\odot ^{-1}$, the average value they obtain for SF galaxies. Note that \citet{Scoville2016} used \alphaco\ =  6.5 \alphacounits\ to derive the molecular gas mass. The more recent exhaustive analysis of \citep{Dunne2022} finds that SF galaxies have a median $\alpha_{\rm 850}$ =5.9, coincidentally close to the value found by \cite{Scoville2016} given that the latter recommend the use of  \alphaco\ = 4.3. Similar to the test of a stacked-derived SFR versus the true mean of individual SFRs described above, we compare these two cases for dust masses and $M_{\textup{mol}}$ masses in Appendix \ref{appendix:sfr} and Fig. \ref{fig:sfr-z-no550} (middle and bottom panels), and reach similar conclusions i.e. we do not expect significant differences between deriving quantities from a stacked image, as compared to averaging individually measured values.

For individually detected or stacking detected CO emission lines, we convert the emission line flux into a molecular gas mass following \cite{Solomon1992}, after using CO ladder luminosity ratios from \cite{Carilli2013} to convert luminosities of higher CO rotational transitions into CO ($J=1\to0$) luminosities, and \alphaco\ = 4.3 \alphacounits\ \citep{Dunne2022}. 

\section{Results}
\label{sec:results}

Our stacking codes provide mean and median stacked results (for both continuum and spectral stacking). We performed the spectral stacking using a stamp-size of 
1\arcsec\ and the continuum stacking with a stamp-size of 20\arcsec $\times$ 20\arcsec.
The continuum stacking code also provides separate results for the full sample, only individually detected sources, and only individually undetected sources. 
Even though the number of individual continuum detections in the ALCS fields is relatively low (e.g. 95 of the SNR $>$5 continuum detections in the ALCS fields  \citep{Fujimoto2023} are in the \kb\ catalogues), the crowded and clustered nature of our source catalogue means that a few bright individually detected sources could bias both the detected central flux, and the outskirts of the stacked maps. 
We thus, unless otherwise mentioned, use and present  results based on mean continuum stacking after eliminating stamps in which a central or peripheral source is individually detected,  with peak flux $> 5$ times the rms of the stamp. For spectral stacking we also show mean stacking results, unless otherwise stated.
All stacks were performed without weighting.
\begin{figure*}
    \centering
    \begin{subfigure}[t]{0.48\textwidth}
        \centering
        \includegraphics[width =\linewidth]{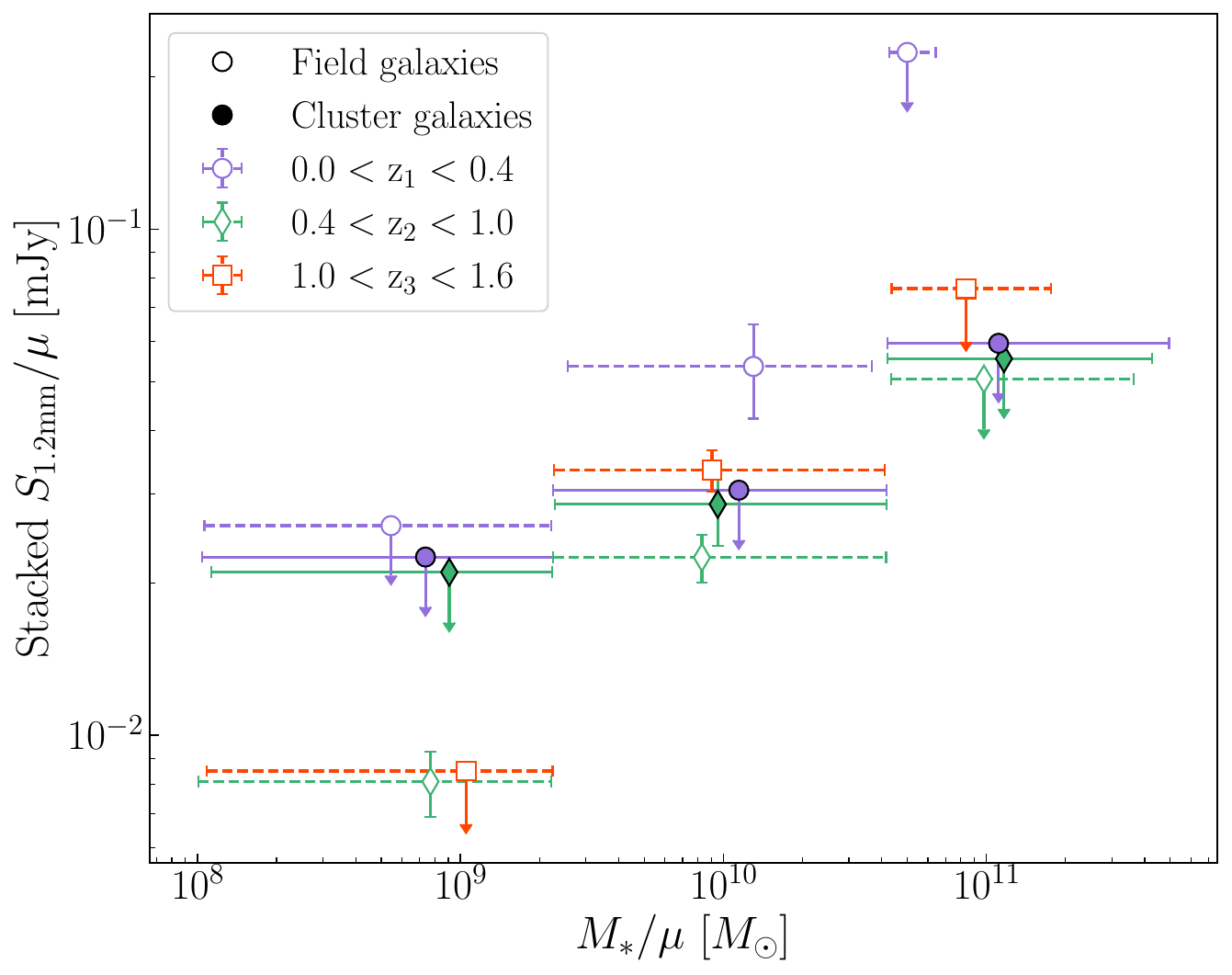}
    \end{subfigure}
    ~
    \begin{subfigure}[t]{0.48\textwidth}
        \centering
        \includegraphics[width =\linewidth]{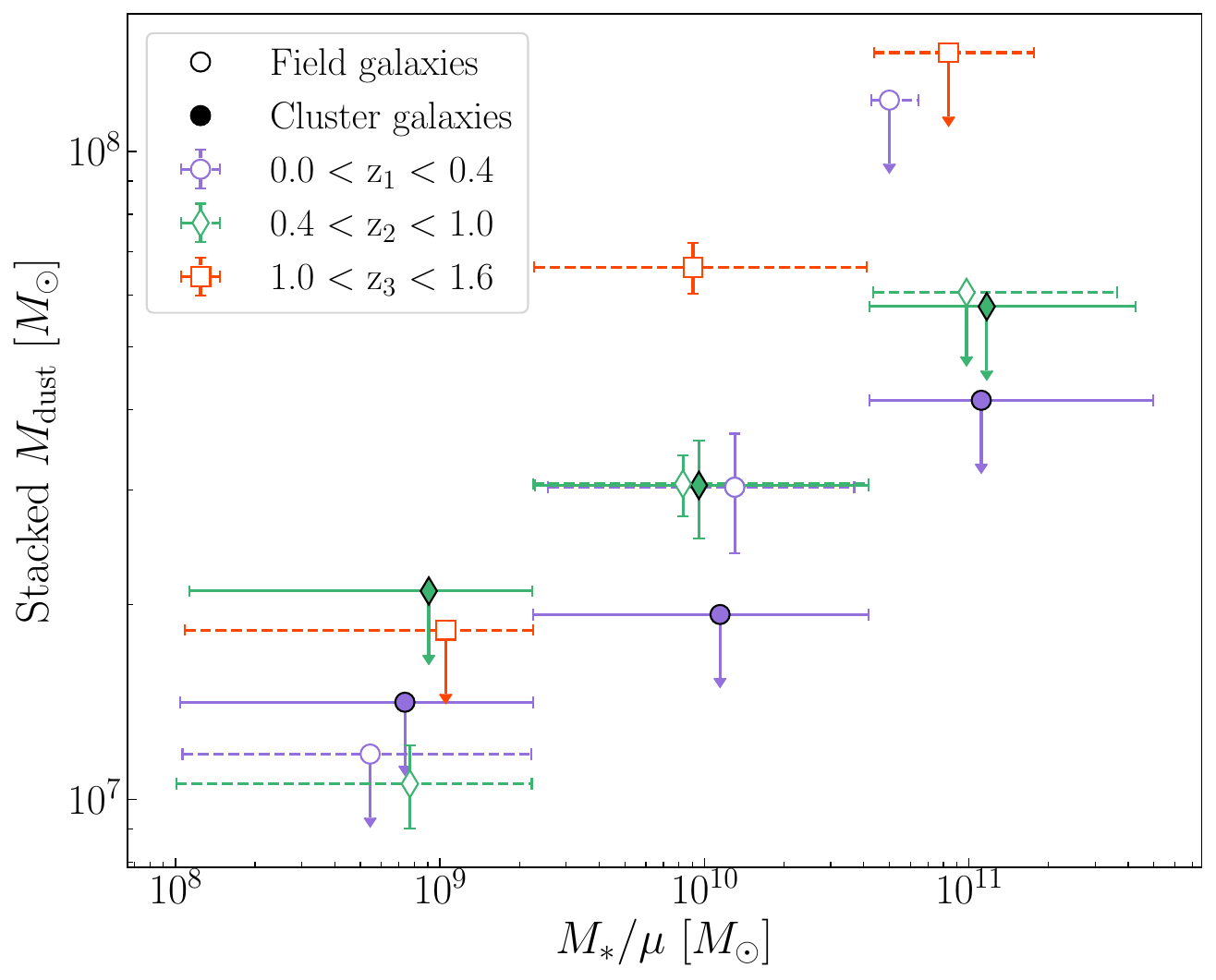}
    \end{subfigure}   
    \caption{Left: the 1.2mm flux in the mean-stacked continuum map of all individually non-detected sources, as a function of the average \kb\ catalogue stellar mass (demagnified), for each redshift bin. 
    Cluster (field) galaxy stacks are shown with filled (open) symbols. Symbols and colours distinguish each redshift bin: purple circles for $0< z_1\leq0.4$, green diamonds  for  $0.4< z_2\leq1.0$ and red squares for  $1.0 < z_3 \leq 1.6$.
    One sigma errors of the integrated fluxes reported by \textsc{Blobcat} are shown (for details see Section \ref{sect:flux_dust_ISM}), and downward pointing arrows typically denote (5$\sigma$) upper limits (see text).  The  horizontal `error bars' of each symbol denote the stellar mass range of galaxies in that bin. 
     Right: as in the left panel but for the 1.2 mm stack-derived dust masses.}
    \label{fig:alma-flux-mdust}
\end{figure*}

\subsection{Full sample stacks}

To illustrate the power, and reliability, of stacking in the ALCS, we stacked all sources - cluster and field - in our ALCS continuum maps using the \kb\ catalogues with the extra filters described in Sect.~\ref{sec:data-cat}, but for the full redshift range of the catalogue, i.e. $0 < z < 12$. The left panel of Fig. \ref{fig:all-spectral-continuum-stacking}  shows the continuum stacked maps of the individually non-detected sources (mean and median stacks).
As expected, the rms of the stacked image decreases by $\sim$$\sqrt{N}$, where N is the number of objects stacked. The mean stacked continuum map of the individual non-detections has an rms of $\sigma_{stack} = 1.4 \mu$Jy: equivalent to a $\sim$8 day integration with ALMA at 250 GHz.

An example spectral stacking result of all galaxies with accurate spectroscopic redshifts is shown in the right panel of Fig. \ref{fig:all-spectral-continuum-stacking}; this figure shows an excerpt of the spectral stack of all sources in which no line was individually detected. While there is no clear line detection here (further below we present a potential CO ($J=4\to3$) stacked line detection in a high-SFR subsample), the figure demonstrates the power of spectral stacking in ALMA datasets. 

\subsection{Spectral stacking in subsamples}
We performed spectral stacking in several subsamples within the compiled spectroscopic catalogue. Here we present results for the two subsamples in which stacked emission lines are detected. Note that, except for the results in Sect. \ref{sect:specdetco}, subsamples were chosen by stellar masses and SFRs from the \kb\ catalogues; i.e. derived using photometric, rather than spectroscopic,  redshifts.

\subsubsection{Stack of detected CO ($J=3\to2$) emission lines}
\label{sect:specdetco}
A stacked spectrum of all 5 sources for which redshifts exist in our compiled spectroscopic catalogue and in which the CO ($J=3\to2$) line is individually detected, is shown in the left panel of Fig. \ref{fig:spectral-stack-subsamples}. These lines were detected in galaxies in the clusters Abell 370, Abell 2744 and MACS1931.8-2635.
The stacked emission line profile has broader wings as compared to a single Gaussian. While this could be a sign of outflows, it could be only an effect of redshift errors, so we do not further interpret the wings. The best-fit  Gaussian to the mean stacked spectrum gives a peak flux of 8.5 $\pm 0.1$ mJy, total flux of 2.84 $\pm$ 0.05 Jy km/s, and FWHM of 293 $\pm 4$ km/s.
These five galaxies have $z_{\mathrm{spec}} =$ 0.293--0.359, and in the \kb\ catalogues, a log $M_{\ast} = $ 10--10.4 \msun\ and SFR $=$ 1--10.7 \msun/yr, i.e. they are between normal star forming galaxies and Luminous Infrared Galaxies (LIRGs). Thus to convert the CO ($J=3\to2$) flux to a molecular gas mass we use 
(a) the Milky Way (MW) value of
L$^\prime_{\rm CO\ J=3\to2}$/L$^\prime_{\rm CO\ J=1\to0}$ = 0.27 
\citep{Weiss2005,Carilli2013}; note that \citet{Daddi2015} found that the CO SLED of redshift $z \sim 1.5$ normal star forming galaxies is similar to the MW up to CO ($J=3\to2$);
(b) normalize to an $\alpha_{\rm{CO}}$ = 4.3 M$_{\odot}$ (K km/s pc$^2$)$^{-1}$ \citep{Dunne2022}. Here $\alpha_{\rm{CO}}$ is the factor used to convert the CO ($J=1\to0$) `surface brightness' luminosity (L$^\prime_{\rm CO\ J=1\to0}$) to the total (molecular hydrogen plus helium) mass in a giant molecular cloud, and could be as low as 1.8 for SMG-like CO SLEDs and 0.4 for ULIRG-like CO SLEDS \citep{Carilli2013}. The implied average molecular gas mass is M$_{\textup{mol}} = 2.3 \times 10^{10} M_{\odot}$ ($\alpha_{\rm{CO}}$/4.3). Using the mean values of the SFRs and stellar masses of these five galaxies from the \kb\ catalogue, the implied average $\mathrm{SFE}$ ($\mathrm{SFE} =$ SFR / M$_{\rm{mol}})$ is $\mathrm{SFE} = 7.9 \times 10^{-10}$yr$^{-1}$ (4.3/$\alpha_{\rm{CO}}$), which is equivalent to a depletion time of $\tau_{\mathrm{dep}} = 1.3$ Gyr. The average molecular gas mass to average stellar mass ratio is thus 1.5 ($\alpha_{\rm{CO}}$/4.3).
The molecular gas mass and depletion time are consistent with previous estimated values for individually detected brightest cluster galaxies (BCGs)  in  galaxy clusters within $z \sim 0.2-0.6$ (e.g. \citealt{Castignani2020}). However, the molecular gas to stellar mass ratio  of 1.5 is relatively high; lowering this ratio to  a more typically determined value would require lowering $\alpha_{\rm{CO}}$, i.e. using a CO SLED closer to those of SMGs or ULIRGs.

\begin{figure*}
    \centering
    \begin{subfigure}[t]{0.48\textwidth}
        \centering
        \includegraphics[width =\linewidth]{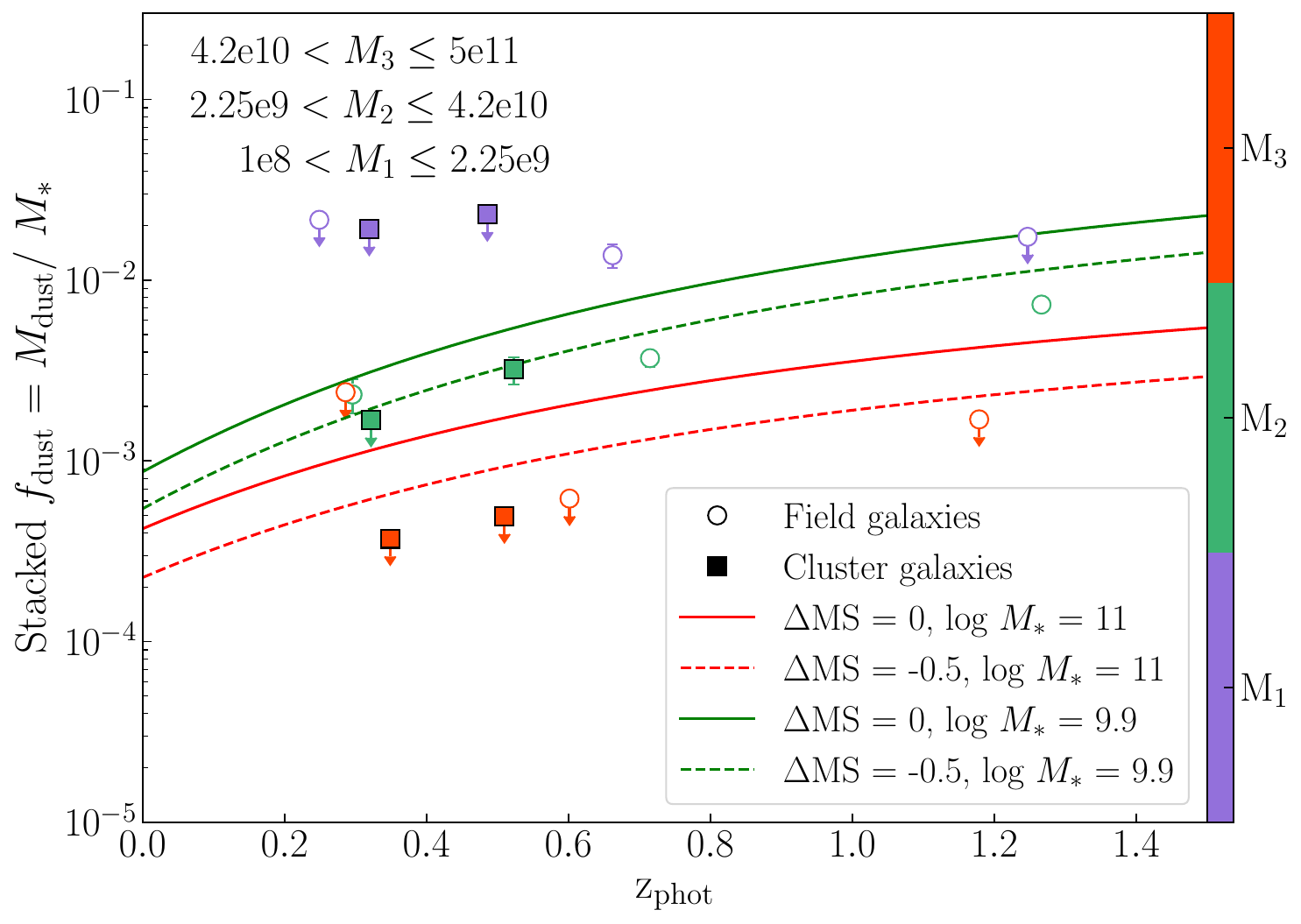}
    \end{subfigure}
    ~
    \begin{subfigure}[t]{0.48\textwidth}
        \centering
        \includegraphics[width =\linewidth]{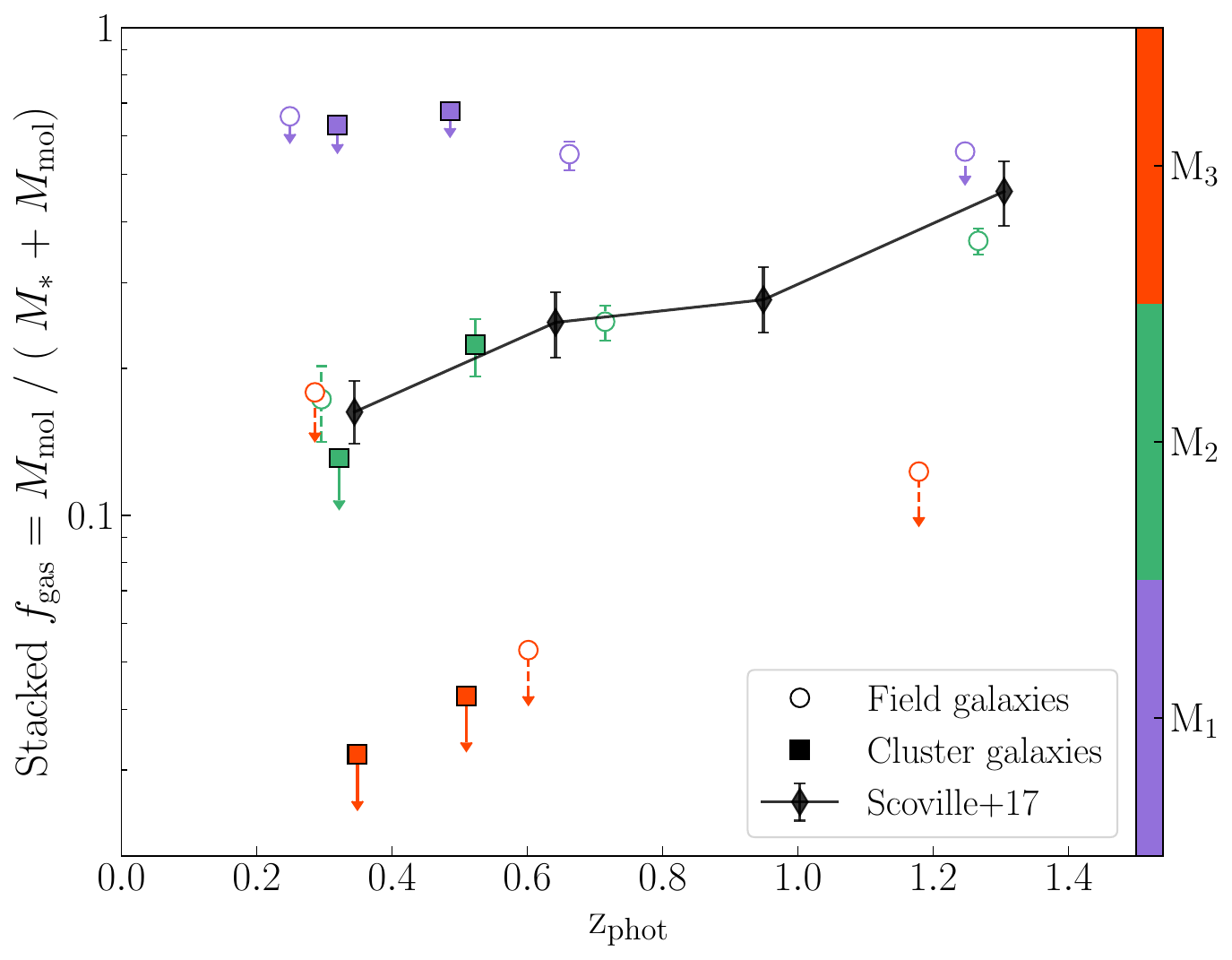}
    \end{subfigure}   
\caption{Left: The stacked dust to stellar mass ratio $f_{\rm{dust}}$ = $M_{\mathrm{dust}}$/ $M_{*}$ (M$_{\rm dust}$ from the stacked 1.2 mm flux and the mean $M_{*}$ (demagnified) from the \kb\ catalogues) as a function of redshift for all redshift and stellar mass bins and for both cluster (squares) and field (circles) galaxies.
Colours distinguish stellar mass bins: purple, green, and red symbols are used for the lowest, middle, and highest stellar mass bins. Downward pointing triangles denote upper limits. The overlaid lines show the expected values \citep[][using a dust to gas ratio of 0.01]{liuet19} for galaxies in the MS (solid; SFR = SFR$_{\rm MS}$) and 0.5 dex below the MS (dashed; SFR $\sim 0.3$ $\times$ SFR$_{\rm MS}$) for galaxies with log $M_{*}$ = 11.0 (red) and 9.9 (green), which correspond to the bin midpoints of our middle and high stellar mass bins.  
Right: As in the left panel, but for the ISM mass fraction (ISM mass to stellar plus ISM mass ratio) as a function of redshift. 
Black diamonds and their error bars show the gas mass fractions derived by \protect\cite{Scoville2017} for star-forming log $M_*$ $\sim$ 11 galaxies in the COSMOS field; these points are connected for easier visualisation.}
    \label{fig:mdust-ismfrac}
\end{figure*}

\subsubsection{Stack of galaxies with high SFRs}
A spectral stack of all `ALCS-SF' galaxies with spectroscopic redshifts which do not have individually detected emission lines did not reveal any detection of a stacked emission line (Fig. \ref{fig:all-spectral-continuum-stacking}).  We instead stacked a subsample of galaxies with the highest \kb\ catalogue SFRs, and thus the highest expected molecular gas masses. To do this we cross matched our combined spectroscopic catalogue with the  combined \kb\ photometric catalogue (for both cluster and field galaxies), and used the spectroscopic redshifts of the former and the SFR estimates of the latter. In this case, we did not use the filters described in Sect. \ref{sec:data-cat}, since in this specific stacking analysis we do not use the photometric redshift, and our results are not thus strongly affected by SFR errors.

We then selected all sources in the uppermost 15-percentile of SFRs, with no individual line detections.  This subsample includes $\sim$300 galaxies with SFRs of $\sim$0.5 to $\sim$2000 \msun/yr.
From that selection, only $\sim$20 of the galaxies had a rest-frame spectrum  between $\sim$460 GHz and $\sim$463GHz, thus contributing to the line, with SFRs of $\sim$1 to $\sim$25 \msun/yr, and a mean of  $\sim$6 \msun/yr. Also, they have stellar masses of log $M_{\ast} = $ 8.3--10.9 \msun\ and a mean of log $M_{\ast} = $ 10.1 \msun. The spectral stack of these sources resulted in an emission line at the expected location of the CO ($J=4\to3$) line detected with SNR = 4. This emission line is shown in the right panel of Fig. \ref{fig:spectral-stack-subsamples}.

The best-fitting Gaussian to the mean stacked spectrum gives a peak flux, total flux, and width of $0.22 \pm 0.02$ mJy, 45 $\pm$ 6 mJy km/s and $179 \pm 20$ km/s, respectively. Since some of the galaxies in the stack were magnified, we divided this total flux by the mean magnification $\mu$, which corresponds to $\sim$3.6. This yields a demagnified total flux of 12.3 $\pm$ 1.6 mJy.

Following the same procedure as for the spectral stack of the individually detected emission line galaxies above, the implied mean molecular gas mass, using the MW value of
L$^\prime_{\rm CO\ (J=4\to3)}$/L$^\prime_{\rm CO\ (J=1\to0)}$ = 0.17
\citep{Weiss2005,Carilli2013,Daddi2015}, is $M_{\rm{mol}} = 5.9 \times 10^{8} M_{\odot}$ ($\alpha_{\rm{CO}}$/4.3). Using the mean SFRs and stellar masses of the stacked galaxies from the \kb\ catalogues, this implies an average $\mathrm{SFE} = 1.2 \times 10^{-8}$yr$^{-1}$ (4.3/$\alpha_{\rm{CO}}$), equivalent to $\tau_{\mathrm{dep}} = 0.09$ Gyr and an average molecular gas to stellar mass ratio of 0.04 ($\alpha_{\rm{CO}}$/4.3). 
These molecular gas mass, gas fractions and depletion times are consistent with the sample of passive and star-forming local galaxies from \cite{Saintonge2017}.

Of the 20 galaxies which contributed to the stacked CO ($J=4\to3$) emission line (bottom right panel of Fig. \ref{fig:spectral-stack-subsamples}), one galaxy from the cluster MACS0429.6-0253 has a potential individual detection of the CO ($J=4\to3$) line, which was not identified in previous analyses. Eliminating this galaxy from the stack weakens the CO ($J=4\to3$) stacked detection to SNR $\sim 3.3$. In this case the best-fitting Gaussian to the mean stacked spectrum gives a peak flux, total flux, and width of $0.2 \pm 0.02$ mJy, 44 mJy km/s and $195 \pm 0.05$ km/s respectively. Although the fitted parameters of the emission line do not change significantly, the SNR is lowered primarily due to the increase in rms.

If we consider this latter line profile as a non-detection, the $3\sigma$ upper limits to the M$_{\rm{mol}}$ and molecular gas to stellar mass ratio are factor 2 higher than the values listed above,  and the SFE lower limit is half the value of the SFE listed above.

\subsection{Continuum stacking in subsamples}
Similarly to spectral stacking, continuum stacking was performed in different subsamples. Since the \kb\ catalogues provide physical properties, the galaxies in the catalogues were stacked in bins of redshift and (de-lensed) stellar mass. From these bins several physical properties are derived, including dust, gas masses, and SFRs.

\begin{figure*}
    \centering
    \includegraphics[width=\linewidth]{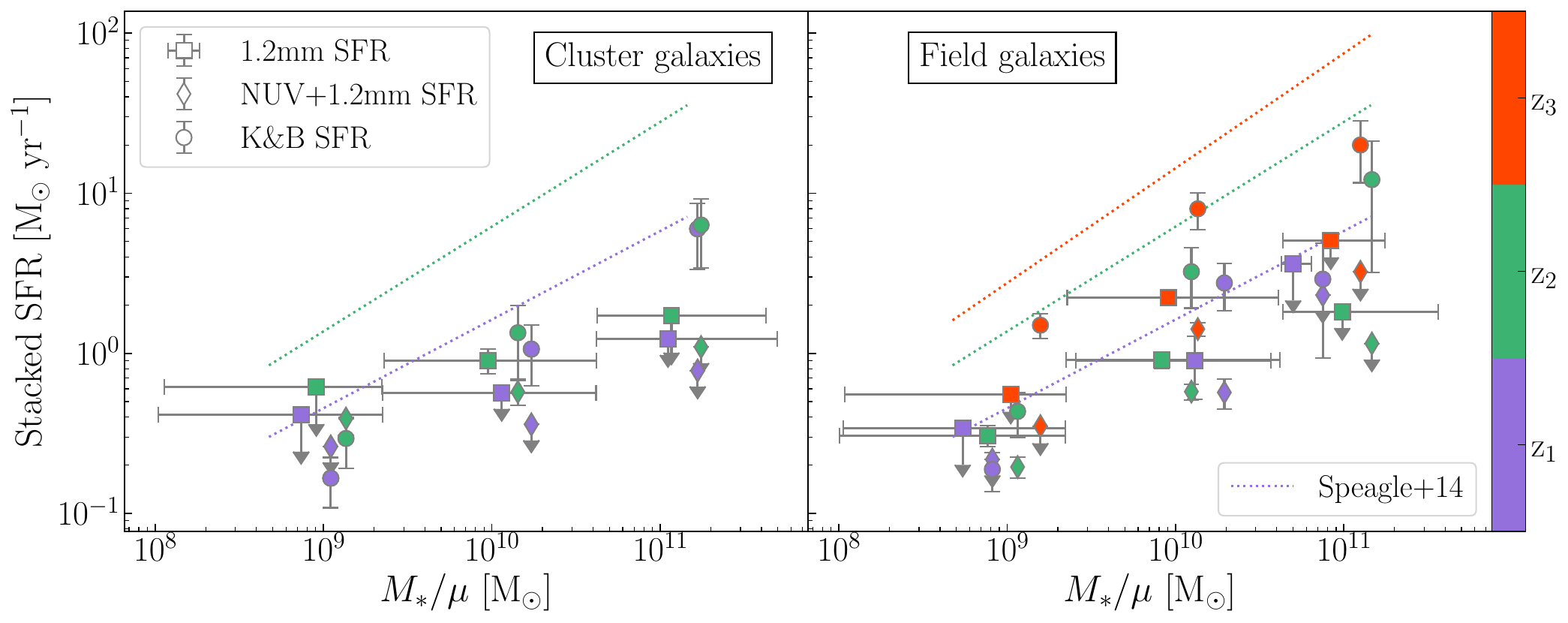}
\caption{Star formation rate as a function of stellar mass bin for cluster (left) and field (right) galaxies using different SFR indicators. Circles denote the SFRs (demagnified) listed in the \kb\ catalogue (from UV to IR spectral fitting), diamonds denote SFRs derived from the combined NUV+1.2mm fluxes (unobscured-SFR), squares show the 1.2mm stacked ALMA derived SFRs (obscured-SFR). Symbols are coloured by the redshift bin following the colour bar on the right. The dotted lines, in the corresponding colours, show the expected MS of \protect\cite{Speagle2014} for each redshift bin.}
    \label{fig:stacked_nuv_sfr}
\end{figure*}

\subsubsection{Fluxes, dust masses and ISM masses}
\label{sect:flux_dust_ISM}

We divided both the cluster and field catalogues into bins of redshift and then sub-bins of stellar mass. The sizes, and ranges, of the bins were driven by the requirement of having sufficient sources in each bin in order to obtain meaningful stacked detections in a significant number of bins. We also selected only individually undetected galaxies. 
The cluster catalogue was thus divided into two redshift bins, $0.0<z_1\leq0.4$ (633 galaxies) and $0.4< z_2\leq1.0$ (817 galaxies), and the field catalogue was divided into three bins, $0.0< z_1\leq0.4$ (304 galaxies), $0.4<z_2\leq1.0$ (1162 galaxies) and $1.0< z_3\leq1.6$ (486 galaxies). In total, this results in 1450 cluster galaxies and 1952 field galaxies.
The highest redshift field bin was used primarily as a sanity check of our results at lower redshifts. 
Each redshift bin was then sub-divided into three stellar mass bins: $1\times10^{8}<M_{1}\leq2.25\times10^{9}$, $2.25\times10^{9}<M_{2}\leq4.2\times10^{10}$, $4.2\times10^{10}<M_{3}\leq5\times10^{11}$.

Figures \ref{fig:cluster-bins} and \ref{fig:stack-median-field} show the stacked maps obtained for the stacks of all cluster and field galaxies, in the redshift and stellar mass bins mentioned above, respectively.  
In each stacked map, we used the source extraction software \textsc{Blobcat} \citep{Hales2012}, which detects sources via a flood fill algorithm. When a blob is detected, flux densities of Gaussian fits to the blob are provided, along with measures of SNR, peak flux, and spatial position.

We consider a stacked source as a detection if: (a) a blob at SNR $\geq$ 5 is found, whose peak position falls in the central 1/3rd of the stamp; or (b) a blob at SNR $\geq$ 4 is found in the  central 1/3rd of the stamp and  no other \textsc{Blobcat} detection is obtained in the map.

For detected sources, we list the SNR corrected for peak bias and the integrated flux corrected for clean bias (see \cite{Hales2012} for details) reported by \textsc{Blobcat}, together with their associated errors. For non-detections, we calculated the mean rms (standard deviation) of the stacked map after excising the central 1/3rd of the map, and reported a 5$\sigma$ upper limit based on these rms.

The 1.2 mm flux (the total flux of the Gaussian fit) or upper limit for each redshift bin in these stacked maps, as a function of the stellar mass bin, are shown in the left panel of Fig. \ref{fig:alma-flux-mdust}. In order to take into account the flux from lensed galaxies, the stacked flux per bin was divided by the mean lensing magnification factor $\mu$ of the galaxies within that bin.
The derived dust masses from these stacked fluxes are shown in the right panel of Fig. \ref{fig:alma-flux-mdust}.

Since the conversion between the $y$-axes of these two panels is almost linear, the results are similar. For the first redshift bin ($0<z_1\leq0.4$), few conclusions can be drawn since many of the bins are undetected. But it is notable that for field galaxies, dust masses increase from the lowest stellar mass bin ($1\times10^{8}<M_{1}\leq2.25\times10^{9}$) to the middle stellar mass bin ($2.25\times10^{9}<M_{2}\leq4.2\times10^{10}$), by at least factor 2.  
The same pattern is seen for the next redshift bin ($0.4<z_2\leq1.0$), for both cluster and field galaxies (factors $\sim$3 and  $\sim$1.3, respectively), and for the highest (field-only) redshift bin ($1.0<z_3\leq1.6$ ; factor $\sim$4).

Dust masses in cluster galaxies are similar those of field galaxies for the middle redshift bin (green) and the middle stellar mass bin ($2.25\times10^{9}<M_{2}\leq4.2\times10^{10}$), but are higher by $\sim$1.5 for field galaxies at $0<z_1\leq0.4$ in the middle stellar mass bin.
In terms of redshift, dust masses increase by factor $\sim$1.5 in cluster galaxies between $0.0<z_1\leq0.4$ and  $0.4<z_2\leq1.0$, while for field galaxies they increase by factor $\sim$2 between $0.0<z\leq1.0$ and $1.0<z_3\leq1.6$.

The left panel of Figure \ref{fig:mdust-ismfrac} shows the redshift evolution of $f_{\rm{dust}}$ (the dust to stellar mass ratio) and compares them with the expected values from \cite{liuet19}, assuming a dust to gas ratio of 0.01. We show the expected values for MS galaxies of $ \log(M_*/M_\odot) = 11$, and the values for galaxies with a MS offset of 0.5 dex. Since this falls within the range of the highest stellar mass bin, $4.2\times10^{10}<M_{3}\leq5\times10^{11}$, the lines are shown with the same colour as the galaxies in that bin (red). Similarly, the expectations for MS galaxies of  $\log(M_*/M_\odot) = 9.9$ (and the ones with an offset of 0.5 dex from the MS) are shown in green, since they fall within the middle stellar mass bin $2.25\times10^{9}<M_{2}\leq4.2\times10^{10}$.

Since the $M_1$ (purple) and $M_3$ 
(red) stellar mass bins have few to no detections, most conclusions can be drawn only for galaxies in the middle stellar mass bin ($2.25\times10^{9}<M_{2}\leq4.2\times10^{10}$ ; green). For field galaxies in this bin, the  mean $f_{dust}$ appear near the expectation of MS galaxies only for the lowest redshift bin $0.0<z_1\leq0.4$, and then at higher redshifts, the dust fractions fall below these expectations by more than the $0.5$ dex offset shown in dotted lines. Cluster galaxies at $0.4<z_2\leq1.0$ in the middle stellar mass bin are within the $0.5$ dex offset from the expected MS. 
Finally, while the highest stellar mass bin $4.2\times10^{10}<M_{3}\leq5\times10^{11}$ has only upper limits to dust masses, these  mean $f_{dust}$ upper limits still appear below the MS expectations by more than $0.5$ dex. 
This implies that field galaxies at higher redshifts ($0.4<z<1.6$) in the middle mass bins, and both cluster and field galaxies in the high stellar mass bins, are less dusty than expected from MS galaxies.

\begin{figure*}
    \centering
    \begin{subfigure}[t]{0.48\textwidth}
        \centering
        \includegraphics[width =\linewidth]{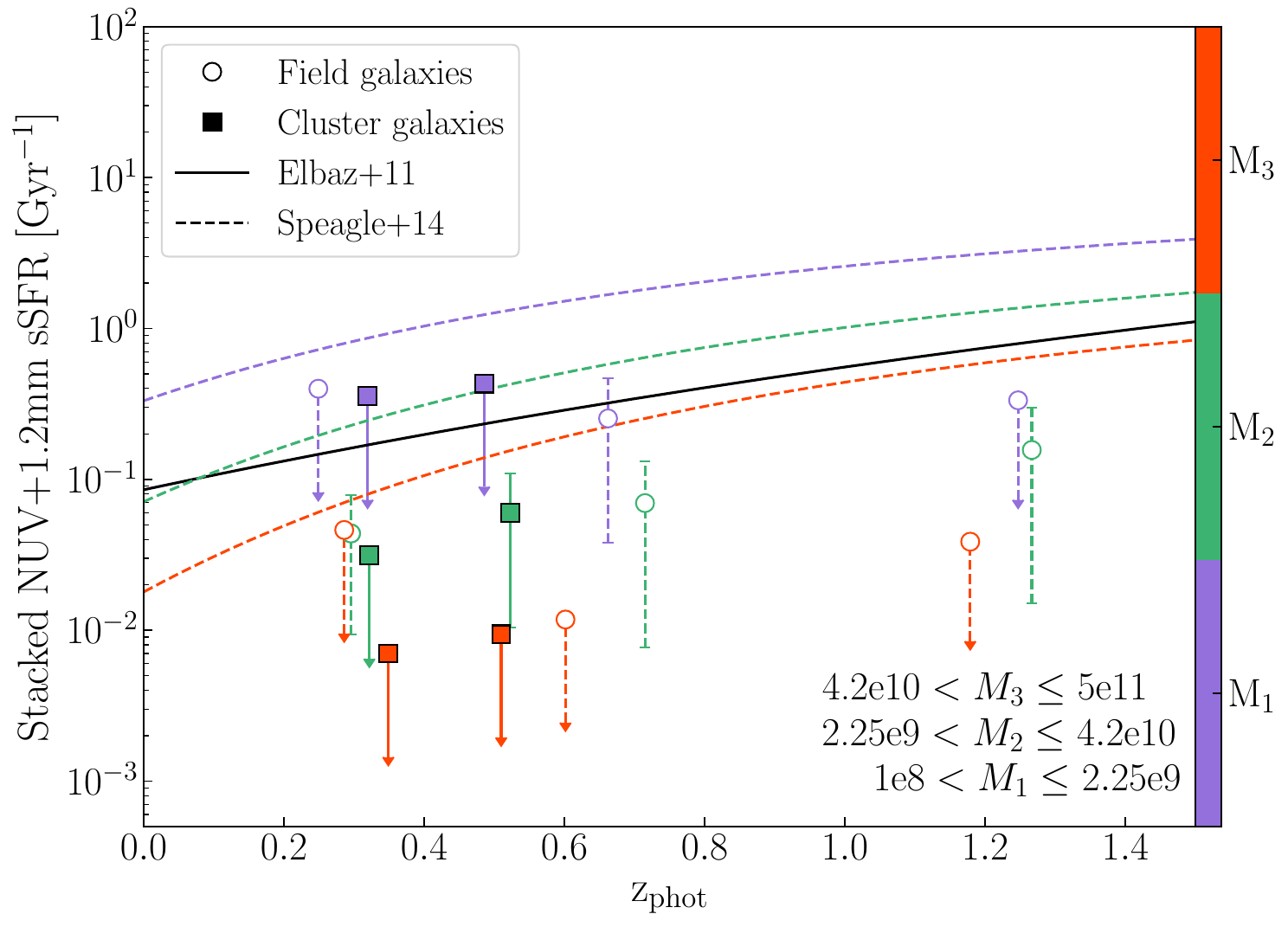}
    \end{subfigure}
    ~
    \begin{subfigure}[t]{0.48\textwidth}
        \centering
        \includegraphics[width =\linewidth]{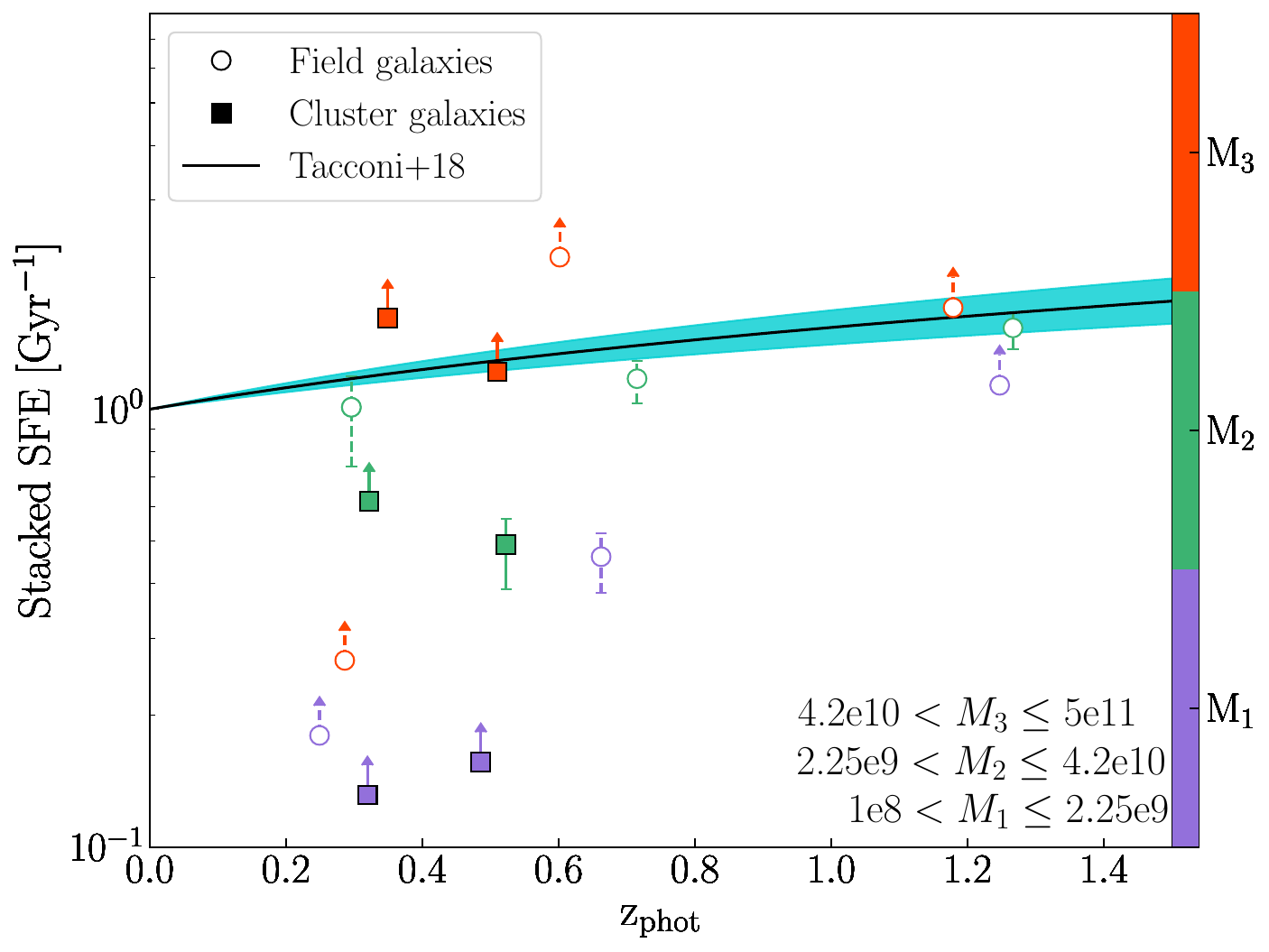}
    \end{subfigure}   
\caption{Specific SFR (sSFR; left panel) and star formation efficiency (SFE; right panel)  as a function of redshift, for cluster (squares) and field (circles) stacks. Colours distinguish stellar mass bins: purple, green, and red symbols are used for the lowest, middle, and highest stellar mass bins. Downward pointing triangles denote upper limits. 
In the left panel, the sSFR is derived using the sum of our stacked SFR and the NUV SFR, and the \kb\ stellar mass. The solid black line shows the \citet{Elbaz2011} main sequence evolution with redshift, and the coloured curves show the MS for each stellar mass bin following \citet{Speagle2014}. 
In the right panel, the SFE is derived from the \kb\ SFR and the 1.2 mm stacked flux derived gas mass, and the  
solid line shows the relationship of SFE with redshift derived by  \protect\cite{Tacconi2018} in the PHIBBS \protect\citep{Tacconi2013} sample (using SFE $\propto$\ \tdep $^{-1}$). The cyan region shows a typical uncertainty of 0.3 dex.}

    \label{fig:ssfr-sfe}
\end{figure*}
 
The right panel of Fig. \ref{fig:mdust-ismfrac} shows the molecular gas fraction ($f_{\rm{gas}}$ = $M_{\rm mol}$/($M_{*}$ + $M_{\rm mol})$) as a function of redshift for the sample bins, and compares these with the fractions seen in log $M_{*}$ $\sim$ 11 galaxies from the COSMOS field \citep{Scoville2017}. It is important to note that though \cite{Scoville2017} refer to measuring ISM  (i.e. molecular plus atomic) masses in their equation 14, this equation is the same as the one used to derive molecular gas masses in \cite{Scoville2016}. 

The gas fraction of the highest stellar mass bins ($4.2\times10^{10}<M_{3}\leq5\times10^{11}$) at all redshifts, and for both cluster and field galaxies (red symbols), are at least $\sim$0.5 to 1 dex lower than the mean values found by  \cite{Scoville2017} in samples of galaxies with similar stellar masses. On the contrary, for the intermediate mass bin $2.25\times10^{9}<M_{2}\leq4.2\times10^{10}$, gas fractions are similar for both cluster and field galaxies, within uncertainties. 

\subsection{Continuum-stacking: SFR, sSFR, and SFE}

The 1.2 mm stacked fluxes shown in the left panel of Fig. \ref{fig:alma-flux-mdust} were converted into SFRs using the mean redshift of galaxies in each redshift and stellar mass bin, and our SFR conversion method. The results are shown in the left (cluster galaxies) and right (field galaxies) panels of Fig. \ref{fig:stacked_nuv_sfr}. This figure shows the 1.2 mm stacked SFRs (squares), the mean SFRs of all galaxies in each bin from the \kb\ photometric catalogues (circle symbols), the corrected SFRs (diamond symbols) derived using the combined NUV and 1.2mm stacked fluxes and the expected MS from \cite{Speagle2014} for each redshift bin. Since \cite{Speagle2014} assumes a Kroupa IMF, we convert it to a Chabrier IMF dividing by 1.06 \citep{Zahid2012}. Since many dots are shown per bin, a small offset in $M_*$ was added per bin for illustrative purposes. 

Overall, we see that the estimates of SFR from \kb\ (circles) are higher that the estimates for NUV+1.2mm SFRs (diamonds) for the middle and high stellar mass bins ($2.25\times10^{9}<M_\odot\leq5\times10^{11}$) at all redshifts in cluster galaxies ($0<z\leq1.0$), for all mass bins ($1\times10^{8}<M_\odot\leq5\times10^{11}$) at the middle and high redshift bins for field galaxies ($0.4<z\leq1.6$), and for the middle mass and low redshift bin. The difference between \kb\ and NUV+1.2mm SFRs in these bins ranges between $\sim$0.4 to $\sim$1 dex for both cluster and field galaxies.
We only see higher/similar (within the uncertainty) NUV+1.2mm SFRs than \kb\ SFRs for the low mass bin in clusters ($0<z\leq1.0$) and for the low redshift bin in field galaxies for $1\times10^{8}<M_{1}\leq2.25\times10^{9}$ and $4.2\times10^{10}<M_{3}\leq5\times10^{11}$.

Similarly, \kb\ estimates (circles) appear higher than 1.2mm stacked highly obscured SFRs (squares), by at least $\sim$0.2 to $\sim$0.7 dex for cluster galaxies, except in the first stellar mass bin in which is not possible to reach a conclusion due to the upper limits. For field galaxies, is not possible to know for the low mass $M_1$ and high mass $M_3$ bins at the first redshift bin due to the upper limits, but for the rest of the bins, \kb\ estimates are higher than 1.2mm SFRs, by at least $\sim$0.2 to $\sim$0.8 dex.

\kb\ SFRs estimates (circles) appear to clearly increase with higher stellar masses for all bins in both cluster and field galaxies, with the only exception being the change between middle ($2.25\times10^{9}<M_{2}\leq4.2\times10^{10}$) to higher ($4.2\times10^{10}<M_{3}\leq5\times10^{11}$) stellar mass in field galaxies, in which the values are similar or higher within the uncertainty. NUV+1.2mm SFRs also seem to be increasing with stellar mass in all bins, with maybe a less noticeable increase in the low ($1\times10^{8}<M_{1}\leq2.25\times10^{9}$) to middle mass bin for cluster galaxies. Lastly, for 1.2mm SFR, it is hard to make conclusions on the evolution of SFR with stellar mass since many of the bins are upper limits.

In general, we see that \kb\ (circles) SFRs estimates are closer to the expected values for MS galaxies from \cite{Speagle2014}, when compared to the other SFRs estimates. The high mass bin at $0<z_1\leq0.4$ for cluster galaxies and the middle mass bin at $0<z_1\leq0.4$ for field galaxies, are shown in agreement with the MS expectations, within uncertainties. The rest of the \kb\ estimates fall below the expected values by $\sim$0.2 to $\sim$0.7 dex. In comparison, the NUV+1.2mm (diamonds) and 1.2mm (squares) SFRs fall below the expectations by at least $\sim$0.2 to $\sim$1.3 dex.

To better visualise the evolution of sSFR with redshift, Fig. \ref{fig:ssfr-sfe} re-plots the data of Fig. \ref{fig:stacked_nuv_sfr}, and compares the data points to the expected evolution of the MS with redshift \citep{Elbaz2011,Speagle2014}. In this case, the sSFR was computed using the combined NUV and 1.2mm SFR.

Overall, the sSFRs of both cluster and field samples appear lower than the MS values at all redshifts. For cluster galaxies, they are lower by at least $\sim$0.2 for $1\times10^{8}<M_{1}\leq2.25\times10^{9}$, $\sim$0.5 for $2.25\times10^{9}<M_{2}\leq4.2\times10^{10}$ and $\sim$0.7 for $4.2\times10^{10}<M_{3}\leq5\times10^{11}$. For field galaxies, they are lower by a range of at least $\sim$0.2-0.6 for $M_{1}$, $\sim$0.5-0.6 for $M_{2}$ and $\sim$0.1-0.8 for $M_{3}$.

The right panel of Fig. \ref{fig:ssfr-sfe} shows the SFE as a function of redshift for our galaxies, and compares these with the results of \citet{Tacconi2018}. The SFE was derived using the SFR from the \kb\ photometric catalogues and the molecular mass derived from the stacked 1.2 mm fluxes.

The only detected bin for cluster galaxies, $2.25\times10^{9}<M_{2}\leq4.2\times10^{10}$ at $0.4<z_2\leq1.0$ shows a SFE is lower than the values of \citet{Tacconi2018} by $\sim$0.7 dex. For field galaxies, we see that the galaxies at $1\times10^{8}<M_{1}\leq2.25\times10^{9}$ and $0.4<z_2\leq1.0$ are below the expectations by $\sim$0.8 dex, however, the middle mass bins (green) follow the expectations within the uncertainties at all redshifts. For the rest of the bins is hard to reach any conclusions, due to the upper limits, but we can say that field galaxies at $4.2\times10^{10}<M_{3}\leq5\times10^{11}$ and $0.4<z_2\leq1.0$, show a SFE higher than expected by $\sim$0.5 dex.

\section{Discussion}
\label{sectdis}

The ALCS data have significant legacy value, and comprehensive spectroscopic redshifts in all 33 cluster fields will greatly enhance the science exploitation of these ALMA data.  
Although we compiled the spectroscopic catalogues for 25 out of 33 clusters, many of these catalogues did not contain enough galaxies to perform a more thorough stacking analysis. In particular, there were not enough galaxies contributing to the most common CO lines (CO $J=4\to3$ or CO $J=3\to2$) at these redshifts to be able to find a stack detection.
Furthermore, the lack of physical properties derived from spectroscopic catalogues prevented us from using better redshift estimates in the interest of understanding galaxy evolution as a function of stellar masses or star formation rates.
Nonetheless, we were able to find a stacked detection in ~20 galaxies with the highest SFRs (SFR from the \kb\ catalogues, thus not tracing highly obscured star formation) of the full sample. The stacked CO ($J=4\to3$) line reveals relatively low gas reservoirs with a molecular gas to stellar mass ratio of $\sim$4\% (or $\lesssim$8\%), comparable with values seen in the sample local galaxies from \cite{Saintonge2017}. In contrast, stacking the individually detected CO lines give very high values of $M_{\rm mol}$/$M_{*}$ ($\sim$150\%).

On another hand, our continuum stacking analysis enables a glimpse into the average properties of faint undetected galaxies for both field and cluster galaxies.

In order to select an unbiased sample, we decided to include only star forming galaxies, by selecting the sources above the MS -- 1 dex line. This way we avoided stacking a significant fraction of passive galaxies. For reference, we compared the stacking for our star-forming sample, `ALCS-SF', with a stacking of the passive galaxies sample (In Fig. \ref{fig:sm-sfr-ssfr} this would be including all galaxies between the red lines in the left panels, i.e. between MS -- 1 dex and MS -- 2 dex). In the stack of the passive galaxies, none of the bins for cluster and field galaxies are detected. Also, we compared full stacks between the `ALCS-SF' and a combined sample of `ALCS-SF' plus the passive sample. For the first, we found a weak detection of SNR $ \sim3$ and 0.005 mJy for the stack of cluster galaxies, and a detection of SNR $ \sim10$ and 0.037 mJy for the stack of field galaxies. For the latter, we found a $5\sigma$ upper limit flux of 0.009 mJy for the stack of cluster galaxies, while for the field stack we find a detection of SNR $ \sim8$ and flux of 0.034 mJy. Therefore, a continuum stacking analysis that included passive galaxies into the sample, would have brought the stacked fluxes down and biassed the results.

Despite the fact that we stacked mainly star-forming galaxies, we found that stacked galaxies fall short on the expected values for MS galaxies for SFRs and sSFRs (e.g. \citealt{Elbaz2011,Speagle2014}) when considering their 1.2mm SFRs. For the UV+IR derived SFRs from \kb\, only a few bins are in agreement. 
Moreover, we did not find a big difference between field and cluster galaxies when it came to dust and gas evolution, but for field galaxies it was certainly easier to find stacked detections, as they seemed to show brighter fluxes in general when compared to the cluster.

\section{Conclusions}

We have completed a stacking analysis within the ALMA Lensing Cluster Survey fields, using new continuum and spectral stacking software, which is made public here.

We performed continuum stacking of the 1.2mm flux for $1450$ cluster and $1952$ field, undetected and star-forming, galaxies at intermediate redshifts, in multiple redshift bins over $z = 0$--1.6, each with three stellar mass bins over log $M_{*}$ [$M_{\odot}$] = 8.0--11.7. We also present spectral stacking of individually detected emission lines and a high SFR-selected subsamples of individually non-detected galaxies.

For the spectral stacking, we find a potential (SNR $\sim 4$) stacked line detection of the CO ($J=4\to3$) emission line among $\sim$20 galaxies with the highest SFR, i.e. galaxies with SFRs of $\sim$1 to $\sim$25 $M_{\odot}$/yr. For this line, and assuming $\alpha_{\rm{CO}}$ = 4.3, we derived an stacked molecular gas mass of $M_{\textup{mol}} = 5.9 \times 10^{8} M_{\odot}$ ($\alpha_{\rm{CO}}$/4.3), stacked $\mathrm{SFE}$ = $1.2 \times 10^{-8}$yr$^{-1}$ (4.3/$\alpha_{\rm{CO}}$) and stacked gas-to-stellar ratio of 0.04. We also report the values if we were to consider this line as a $3\sigma$ upper limit, which corresponds to twice the values listed above for $M_{\textup{mol}}$ and molecular gas to stellar ratio, and a SFE lower limit which is half of the SFE value listed above.

Our continuum stacked 1.2 mm fluxes are used to estimate average dust masses, molecular gas masses and SFRs, allowing us to contrast the average properties of cluster and field galaxies, and their evolution with stellar mass and redshift. The conversion of stacked 1.2 mm fluxes to masses and SFRs assumes a single component dust temperature and grey body spectral index, and is done at the mean redshift of the stacked subsample. 

In general, these stacked galaxies show lower SFR and sSFR when compared to values of MS galaxies from \cite{Elbaz2011} and \cite{Speagle2014}. Also, only the galaxies in the middle mass bin $2.25\times10^{9}<M_{2}\leq4.2\times10^{10}$ showed dust and gas content comparable to MS galaxies from \cite{Scoville2017} and \cite{liuet19}, while galaxies in the higher mass bin $4.2\times10^{10}<M_{3}\leq5\times10^{11}$ also seem to have less dust and gas than MS galaxies. 
Something similar is seen for SFE, in which only field galaxies in $M_2$ agree with values of \cite{Tacconi2018}, while the cluster galaxies fall shorter. For the rest of the bins it is hard to reach a conclusion due to upper limits.

When comparing cluster versus field galaxies, we found that although field galaxies were usually brighter than cluster galaxies when trying to find a stack detection, both seemed to have similar trends when it came to their SFR and dust and gas contents.

While our results require confirmation with more comprehensive and accurate catalogues of redshift and stellar mass for the ALCS clusters, they already find lower average gas mass fractions, dust mass fractions, SFR and sSFRs than those than average values found for individually detected galaxies. 

\section{Acknowledgements}

We thank the anonymous referee for their helpful comments.
We acknowledge funding from the Agencia Nacional de Investigación y Desarrollo (ANID, Chile) programs:
Nucleo Milenio TITANs NCN19$-$058 (AG, NN; 
FONDECYT Regular 1171506 (AG, NN), 1190818 (FEB) and 1200495 (FEB); 
ANID Basal - PFB-06/2007 (NN, FEB, GN), AFB-170002 (AG, NN, FEB, GN) and FB210003 (AG, NN, FEB); 
Millennium Science Initiative - ICN12\_009 (FEB);  
Programa Formación de Capital Humano Avanzado (PFCHA) / Magíster Nacional/2019 - 22191646. We additionally acknowledge support from:
JSPS KAKENHI Grant Number JP17H06130 (K. Kohno);
NAOJ ALMA Scientific Research Grant Number 2017-06B (K. Kohno));
the Swedish Research Council (JBJ, K. Knudsen)
the Knut and Alice Wallenberg Foundation (K. Knudsen);
the NRAO Student Observing Support (SOS) award SOSPA7-022 (FS);
the Kavli Foundation (NL);
a Beatriz Galindo senior fellowship (BG20/00224) from the Ministry of Science and Innovation (DE).

This publication uses data from the ALMA program: ADS/JAO.ALMA\#2018.1.00035.L. ALMA is a partnership of ESO (representing its member states), NSF (USA) and NINS (Japan), together with NRC (Canada), MOST and ASIAA (Taiwan), and KASI (Republic of Korea), in cooperation with the Republic of Chile. The Joint ALMA Observatory is operated by ESO, AUI/NRAO and NAOJ.

\section{Data Availability}
The data underlying this article will be shared on reasonable request to the corresponding author.

\bibliographystyle{mnras}
\bibliography{refer}

\appendix

\section{Stacking software}
\label{sectappendixstack}

The stacking software developed in this work is made public on Github. Here we briefly describe the working of these codes and their inputs and outputs. Both routines can be executed as scripts within CASA or as CASA tasks.

\subsection{Continuum Stacking Code}
\label{sect:app_cont_stack}

The continuum stacking task and script
requires a list of fits maps to use (which can be of varying size and resolution), a coordinate file with the position (right ascension and declination in decimal degrees or radians) of each target, and optionally a weight to be used for this target during stacking. The user also specifies the square \textit{stamp size} in arcseconds to be used for sub-map extraction and stacking.

A  continuum sub-map of \textit{stamp size} is extracted for each source in the coordinate file, and these are saved into a fits cube where each channel of the cube  corresponds to an individual source in the input coordinate catalogue. Each stamp, and final stacked image, is divided into a $3\times3$ grid, and the standard deviation is computed using only the 8 edge `quadrants'. 
This cube is later stacked using both the (weighted) mean and median method.
Three stacks are created in each run:  `all-sources', `only-detection' and `only non-detection'. The first case includes all of the stamps. For the only-detection stack, an algorithm identifies sources individually detected,  according to a signal to noise ratio (SNR) specified as an input by the user: a source is considered detected if its peak flux is higher than the user-defined noise (e.g. above 3$\sigma$). If a detection is found in the central 1/3rd of the map, the stamp will be saved into a separate `only-detections' fits cube, and at the end of the process, a stacked map of the individual detections will be created. For the non-detection stack, the code selects only stamps in which sources are not individually detected (in any part of the stamp). Then, optionally, stamps are rotated in multiples of 90$^{\circ}$ in order to cancel out striping or other correlated noise in the maps. Finally, the stamps are saved into a cube and stacked.  
 
The resulting stacked maps are saved to fits files and informative plots are produced. The plots show the average standard deviation $\bar{\sigma}$ for the stamps, the standard deviation ($\sigma$) of the stacked maps and the number of stacked maps (number of stamps).
The stacked maps are shown with and without smoothing. 

\subsection{Spectral Stacking Code}

The spectral stacking code performs stacking within a list of data cubes provided by the user. The inputs are (a) a coordinate file with the positions (right ascension and declination in decimal degrees or radians), and redshifts of the targets to stack, and an optional column with weights for each source; (b) the datacubes to be used (specified by a fits file name and directory), which can be of different sizes and resolutions, and (c) the \textit{stamp size}, i.e. the width in arcseconds of the square region to extract for stacking.

For each source in the coordinate file, the observed spectrum is extracted from the datacube (over an aperture of \textit{stamp size}), and any continuum flux contribution is subtracted (by subtracting the median of the spectra). Intermediate results of this stage, e.g. the extracted spectra, sub-catalogues of sources for which the meaningful spectra were extracted, and those which do not fall within the ALMA datacubes, are stored as ascii files.

Extracted spectra are then converted to rest-frame (with an optional oversampling factor) and both a median- and mean-stacked spectrum are  produced; intermediate results of this stage, e.g. all rest frame spectra, are stored in ascii files. 
After the stacking is performed, the code generates several plots to better understand and visualise the results. Plots created include individual spectra for each galaxy (observed and rest-frame), the stacked spectrum over the entire frequency range or subplots for each 20 to 50 GHz window, and parts of the stacked spectrum centred on known extragalactic emission lines. 

Since intermediate ascii files are saved, when the stacking code is run again (e.g. when adding an extra cube to a previously stacked list of cubes), the user has the option to perform the extraction and conversion to rest-frame only for the new cube(s), and then combine the new and previous extractions to update the final stacked products. This is useful to stack a large number of cubes and to reduce the time needed to extract the observed spectra.

\section{Commutativity of stacking vs. averaging when deriving submm SFRs, dust and gas masses}
\label{appendix:sfr}

This appendix presents additional information and tests on the conversion of (stacked) observed-frame 1.2mm ALMA fluxes into dust masses, $M_{\rm mol}$ masses, and SFRs. 

\begin{figure}
    \centering
    \begin{subfigure}[t]{0.475\textwidth}
        \centering
        \includegraphics[width = 0.9\columnwidth]{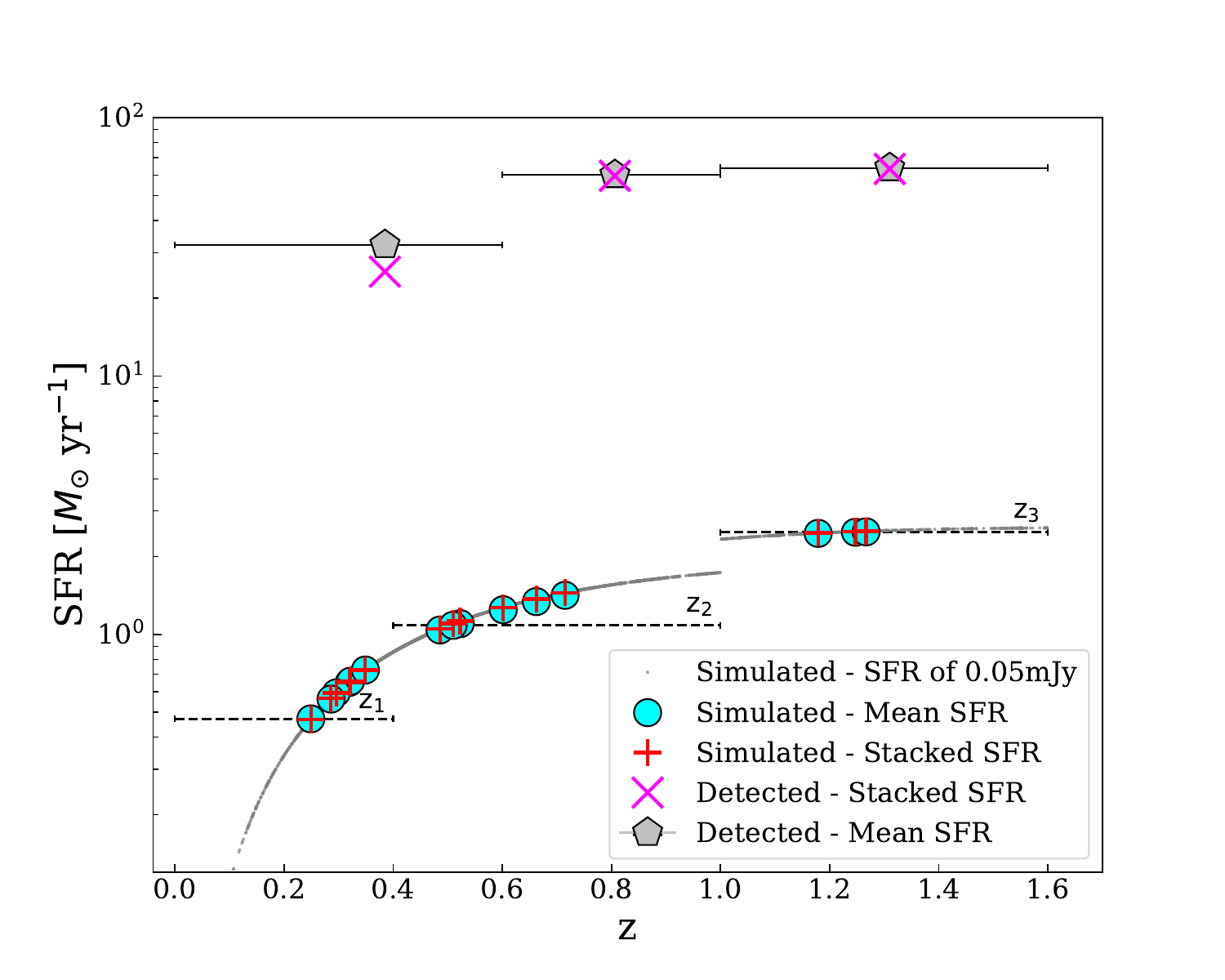}
    \end{subfigure}
\vfill
    \begin{subfigure}[t]{0.475\textwidth}
        \centering
        \includegraphics[width = 0.9\columnwidth]{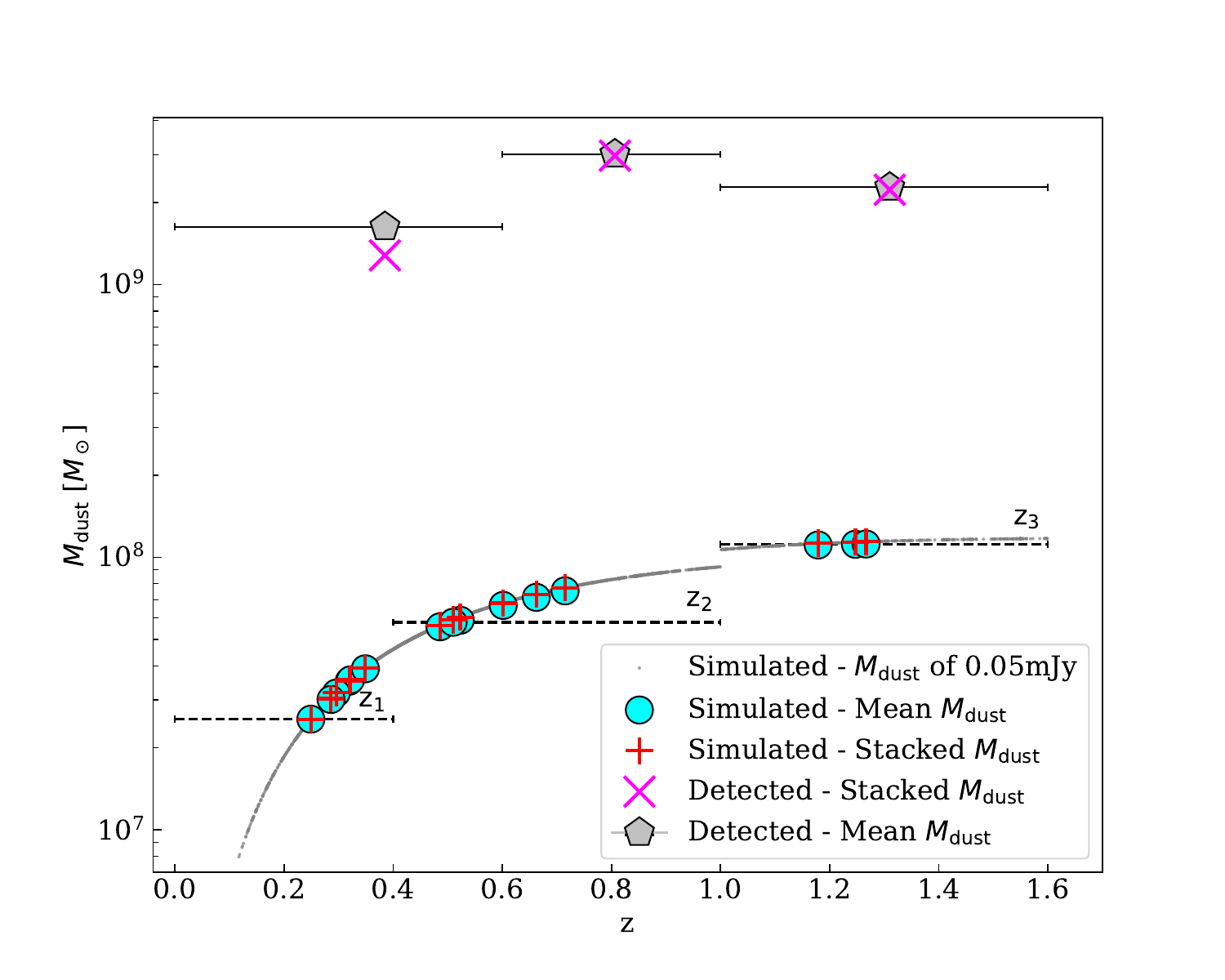}
    \end{subfigure}
\vfill       
    \begin{subfigure}[t]{0.475\textwidth}
        \centering
        \includegraphics[width = 0.9\columnwidth]{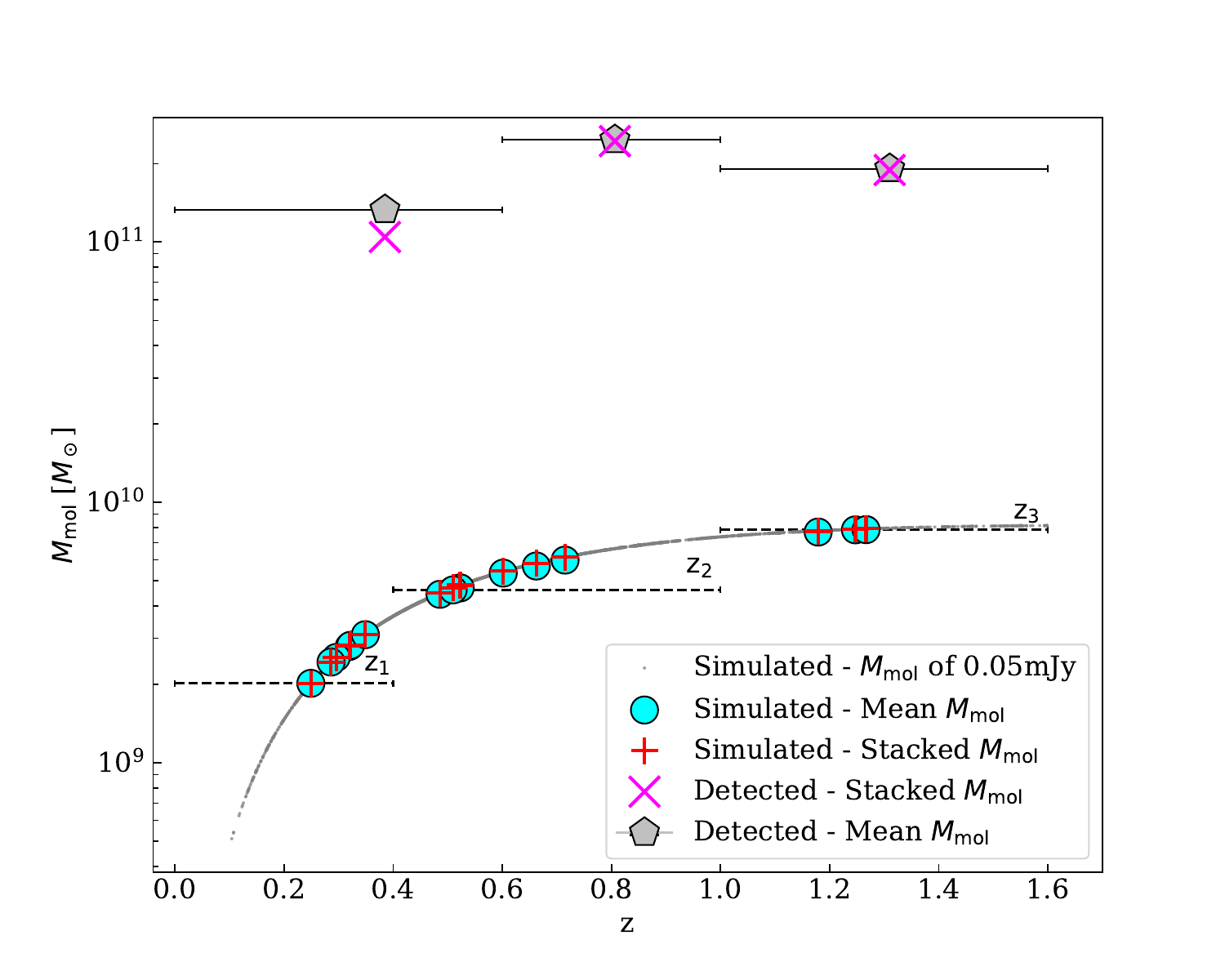}
    \end{subfigure}
\caption{Top panel: SFR for a source with observed 1.2mm ALMA flux of 0.05 mJy, over the redshift range of our sample galaxies (small grey dots). A constant temperature of 22K is assumed at all redshifts. The jump at $z \sim 1$ is due to using equations \ref{eq:Lir1} (\ref{eq:Lir2}) at low (high) redshift. The results of the MC simulations which compare the value derived from the stacked mm flux at the mean redshift of the bin (stacked SFR), with the mean value of SFR of the individual galaxies (true mean SFR; see text), are shown with red plus signs and cyan circles respectively. Dashed black lines show each redshift bin.
Equivalent results for the individually detected mm sources are shown in purple crosses and grey pentagons. The horizontal black errorbars show each redshift bin. Symbols are plotted at the mean redshift of each subsample.
Middle and bottom panel: as in the top panel, but for dust masses and $M_{\rm mol}$ masses.}
    \label{fig:sfr-z-no550}
\end{figure}

We first test for consistency between mean estimates of individual SFRs, dust masses and $M_{\rm mol}$ masses with the values derived from stacking, i.e. the value obtained from stacked fluxes and the mean redshift of the subsample.  
Figure \ref{fig:sfr-z-no550} illustrates the conversion of a constant 1.2mm ALMA flux into SFR (top panel), dust mass (middle panel) and $M_{\rm mol}$ mass (bottom panel)  estimates at different redshifts. Here we use a source with flux 0.05 mJy and black body temperature 22K, over the range of redshifts in our sample. At $z>1$ the estimations are relatively flat, but at $z <1$ the estimated quantity from the stack may not necessarily represent the mean of the individual galaxies. 
We performed MC simulations using the redshift distributions within each of the redshift and stellar mass bins, assigning a uniformly distributed flux, between 0 and 0.1 mJy, to individual galaxies. The reason of this range is that we consider the 2$\sigma$ value of the typical noise of unstacked ALCS continuum maps. Also, this is the reason why we chose to simulate a source of constant flux at 0.05 mJy, which is the mean value. The resulting stacked estimates in each MC iteration were compared to the mean quantity of all galaxies, thus constraining the  error of the former. 
Besides the MC results, we compared the stacked estimates with the mean quantities of individually detected galaxies \citep{Fujimoto2023}.
 As can be seen in the figures both methods give relatively similar results. Thus we feel confident that within a redshift bin we can make meaningful comparisons between the results of each stellar mass bin. 

\section{Continuum stacking maps} \label{sec:cont-images}

This appendix presents the continuum-stacked maps of the subsamples, which are used to derive the observed 1.2 mm flux shown in Fig. \ref{fig:alma-flux-mdust}.

\begin{figure*}
    \centering 
\begin{subfigure}{0.25\textwidth}
 \hspace{0.45\textwidth} \Large  M$_1$
\end{subfigure}\hfil 
\begin{subfigure}{0.25\textwidth}
  \hspace{0.45\textwidth} \Large  M$_2$
\end{subfigure}\hfil
\begin{subfigure}{0.25\textwidth}
  \hspace{0.45\textwidth} \Large  M$_3$
\end{subfigure}\hfil
\medskip

\settoheight{\tempdim}{\includegraphics[width=0.25\textwidth]{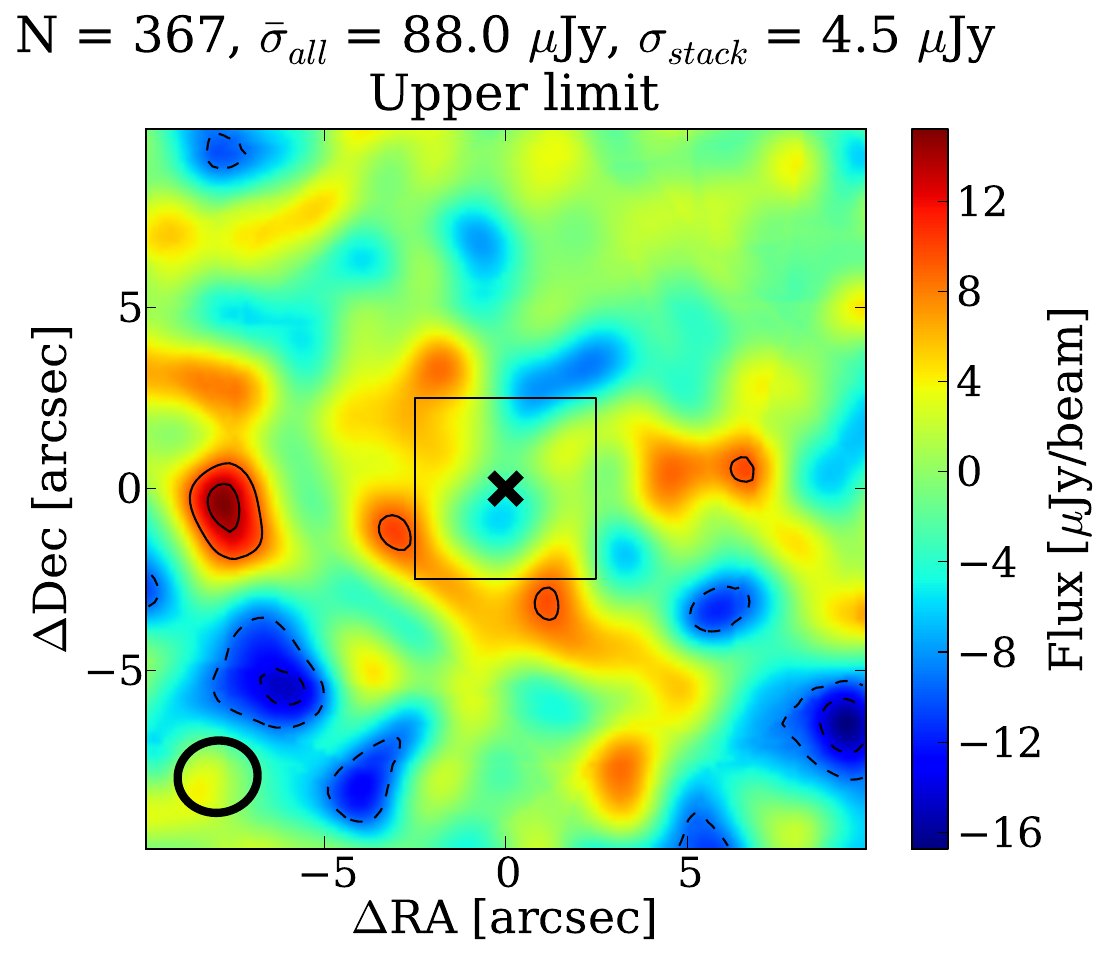}}
\rotatebox{0}{\makebox[\tempdim]{\Large $z_1$}}\hfil
\begin{subfigure}{0.25\textwidth}
  \includegraphics[width=\linewidth]{c_z1m1.pdf}
  \label{fig:1}
\end{subfigure}\hfil 
\begin{subfigure}{0.25\textwidth}
  \includegraphics[width=\linewidth]{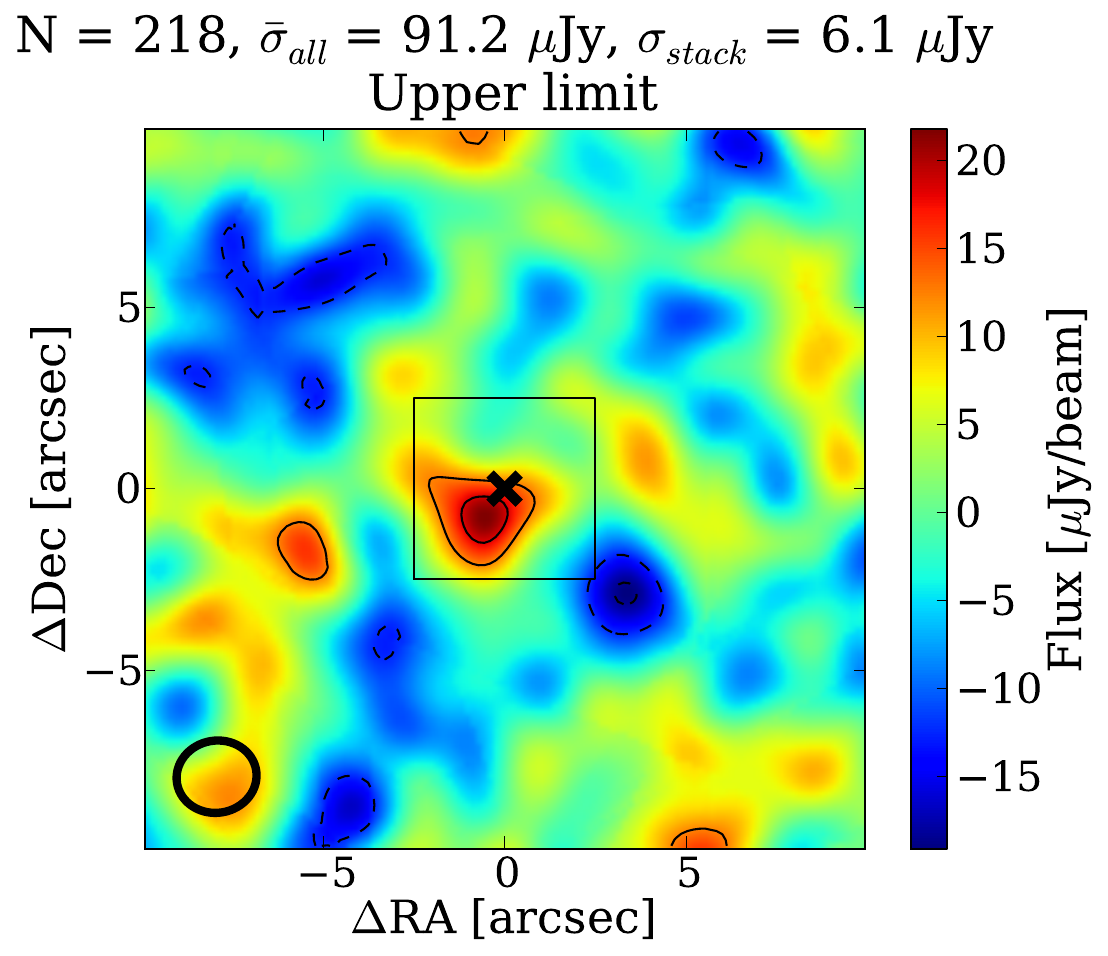}
  \label{fig:2}
\end{subfigure}\hfil
\begin{subfigure}{0.25\textwidth}
  \includegraphics[width=\linewidth]{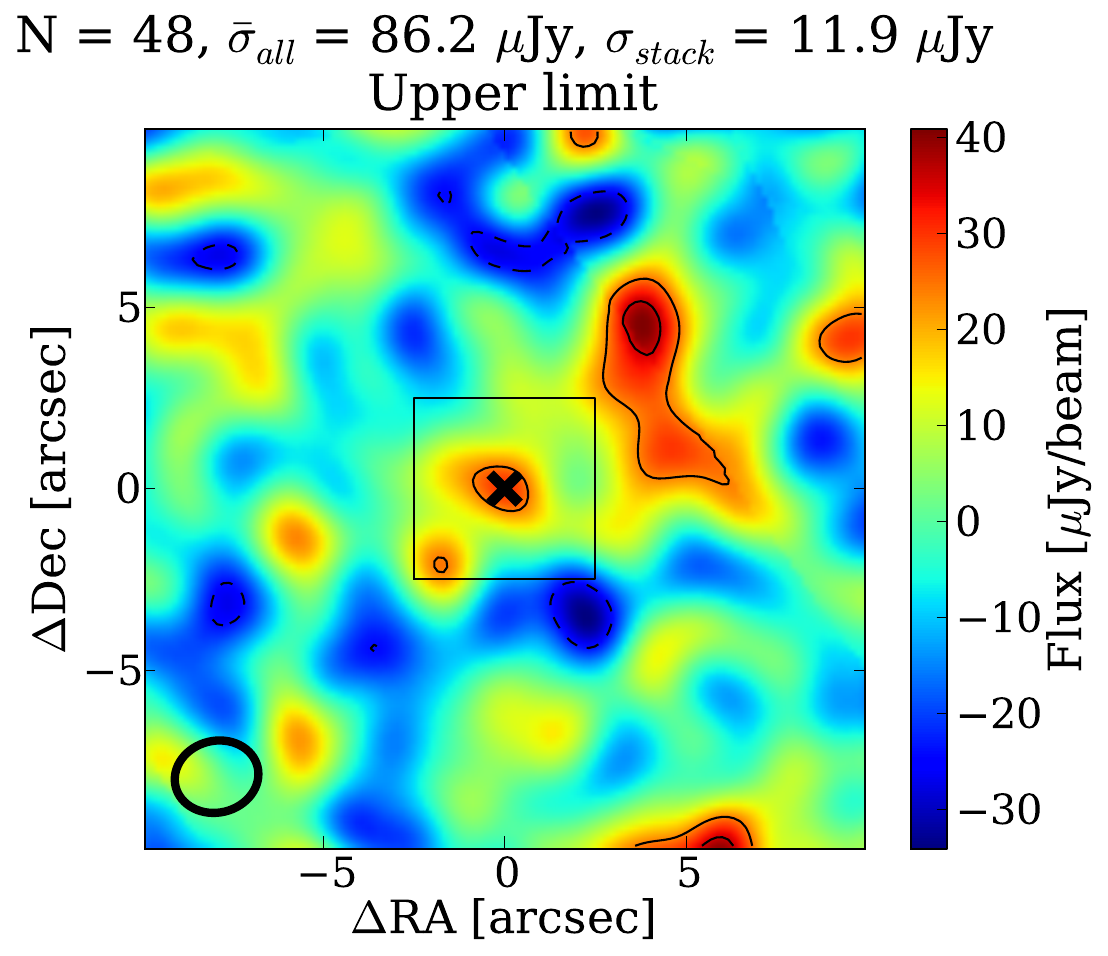}
  \label{fig:3}
\end{subfigure}\hfil
\medskip

\settoheight{\tempdim}{\includegraphics[width=0.25\textwidth]{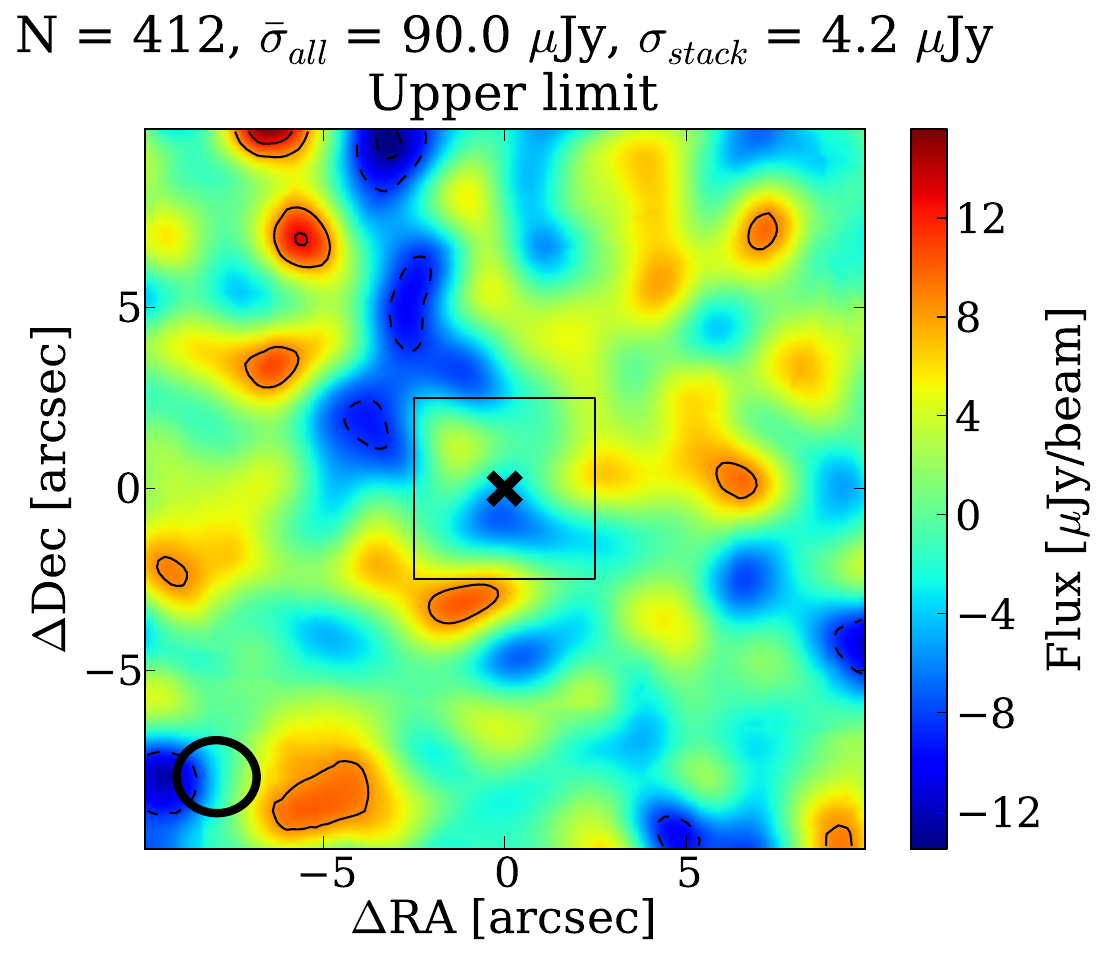}}
\rotatebox{0}{\makebox[\tempdim]{\Large $z_2$}}\hfil
\begin{subfigure}{0.25\textwidth}
  \includegraphics[width=\linewidth]{c_z2m1.pdf}
  \label{fig:4}
\end{subfigure}\hfil 
\begin{subfigure}{0.25\textwidth}
  \includegraphics[width=\linewidth]{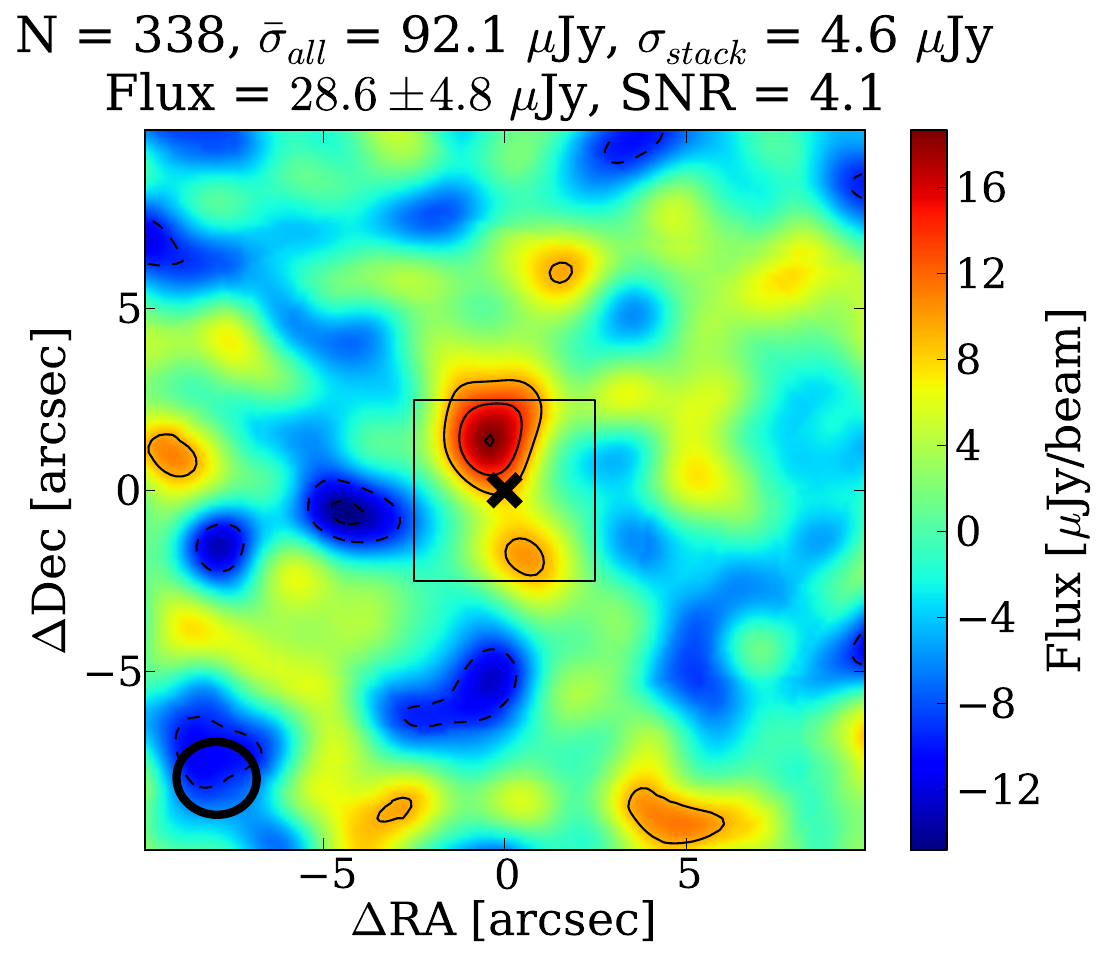}
  \label{fig:5}
\end{subfigure}\hfil  
\begin{subfigure}{0.25\textwidth}
  \includegraphics[width=\linewidth]{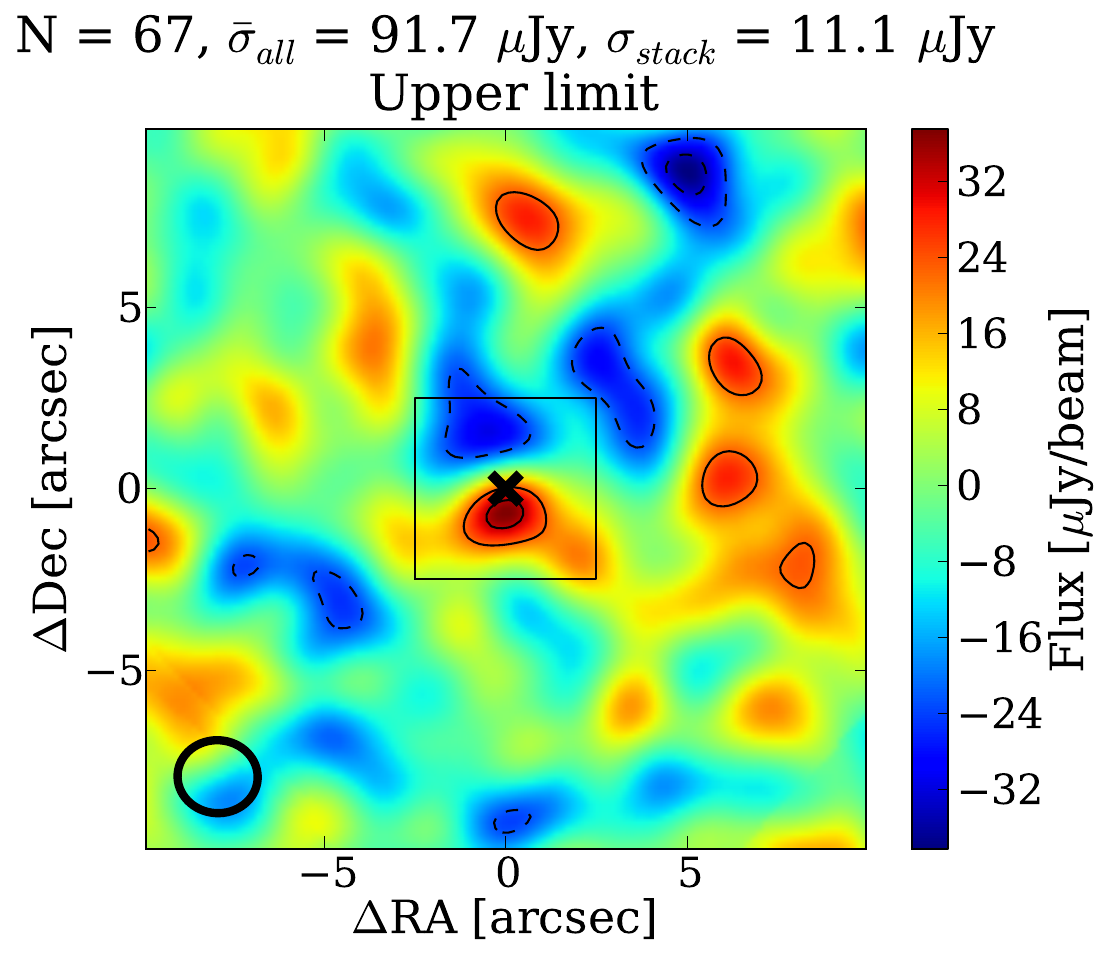}
  \label{fig:6}
\end{subfigure}
\caption{The mean-stacked continuum maps of cluster galaxies, used to derive the fluxes shown in Fig. \ref{fig:alma-flux-mdust}. The top row shows the stacked maps for the redshift bin $0<z_1\leq0.4$ and bottom row shows the stacked maps for the redshift bin $0.4<z_2\leq1.0$. Left panels correspond to the lowest stellar mass bin $1.0\times10^{8}<M_{1}\leq2.25\times10^{9}$,  middle panels to the second stellar mass bin $2.25\times10^{9}<M_{2}\leq4.2\times10^{10}$, and  right panels to the highest stellar mass bin $4.2\times10^{10}<M_{3}\leq5\times10^{11}$.
At the top of each panel, we list the number of maps which entered into the stack (N), the average rms of the input maps  ($\bar{\sigma}_{all}$), the rms of the  stacked map ($\sigma_{stack}$). If the criteria for having a detection in a given stacked map are met (see Section \ref{sect:flux_dust_ISM}), we list the total flux of the fitted Gaussian and SNR (both values from \textsc{Blobcat}), otherwise, we state that the bin is an upper limit. The centre of each stacked image is indicated with a black cross, and the synthesized beam of one of the input images is shown in the bottom left corner. The central 2\farcs5 of the stack image is shown with a black square. Contours are shown at levels of [$-3,-2,2,3,4$] of the rms in the map.}
\label{fig:cluster-bins}
\end{figure*}

\begin{figure*}
    \centering 
\begin{subfigure}{0.25\textwidth}
 \hspace{0.45\textwidth} \Large  M$_1$
\end{subfigure}\hfil 
\begin{subfigure}{0.25\textwidth}
  \hspace{0.45\textwidth} \Large  M$_2$
\end{subfigure}\hfil
\begin{subfigure}{0.25\textwidth}
  \hspace{0.45\textwidth} \Large  M$_3$
\end{subfigure}\hfil
\medskip

\settoheight{\tempdim}{\includegraphics[width=0.25\textwidth]{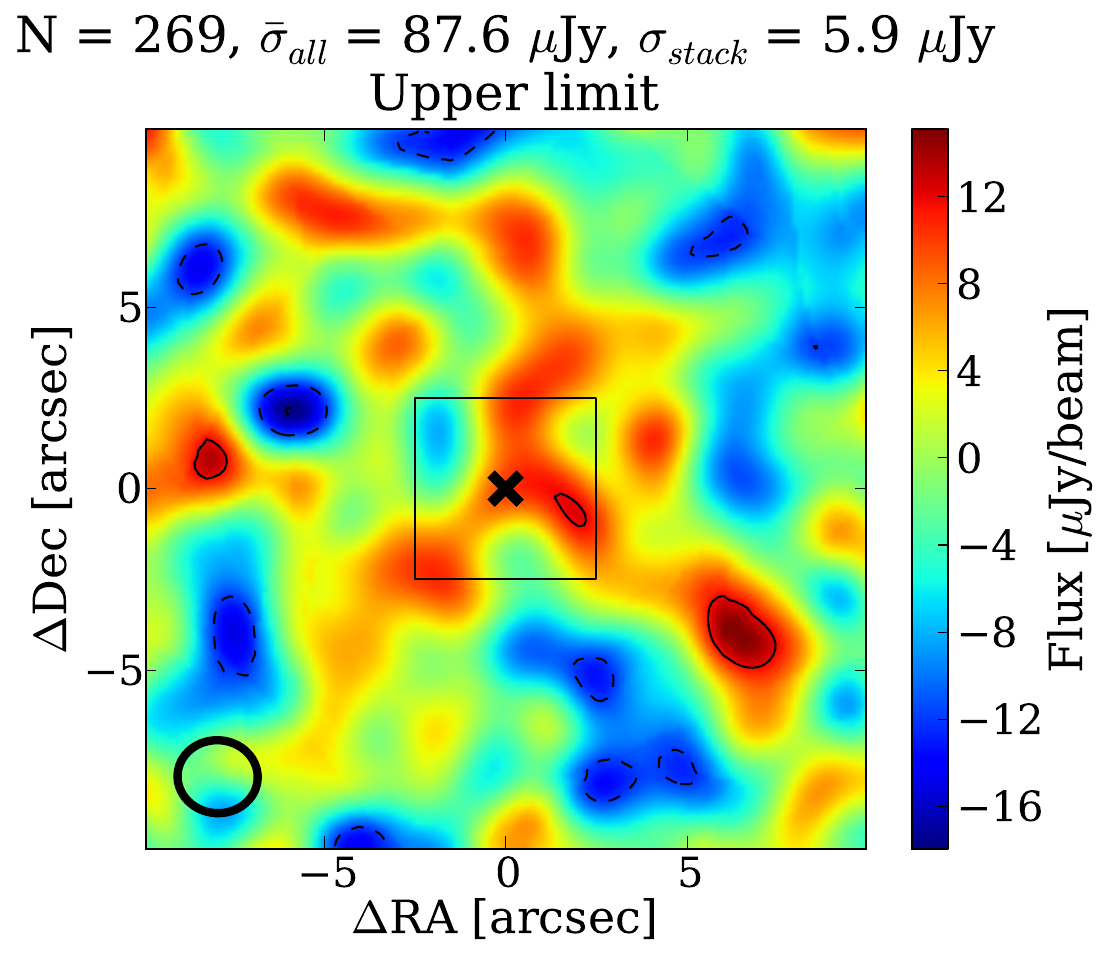}} 
\rotatebox{0}{\makebox[\tempdim]{\Large $z_1$}}\hfil
\begin{subfigure}{0.25\textwidth}
  \includegraphics[width=\linewidth]{f_z1m1.pdf}
\end{subfigure}\hfil 
\begin{subfigure}{0.25\textwidth}
  \includegraphics[width=\linewidth]{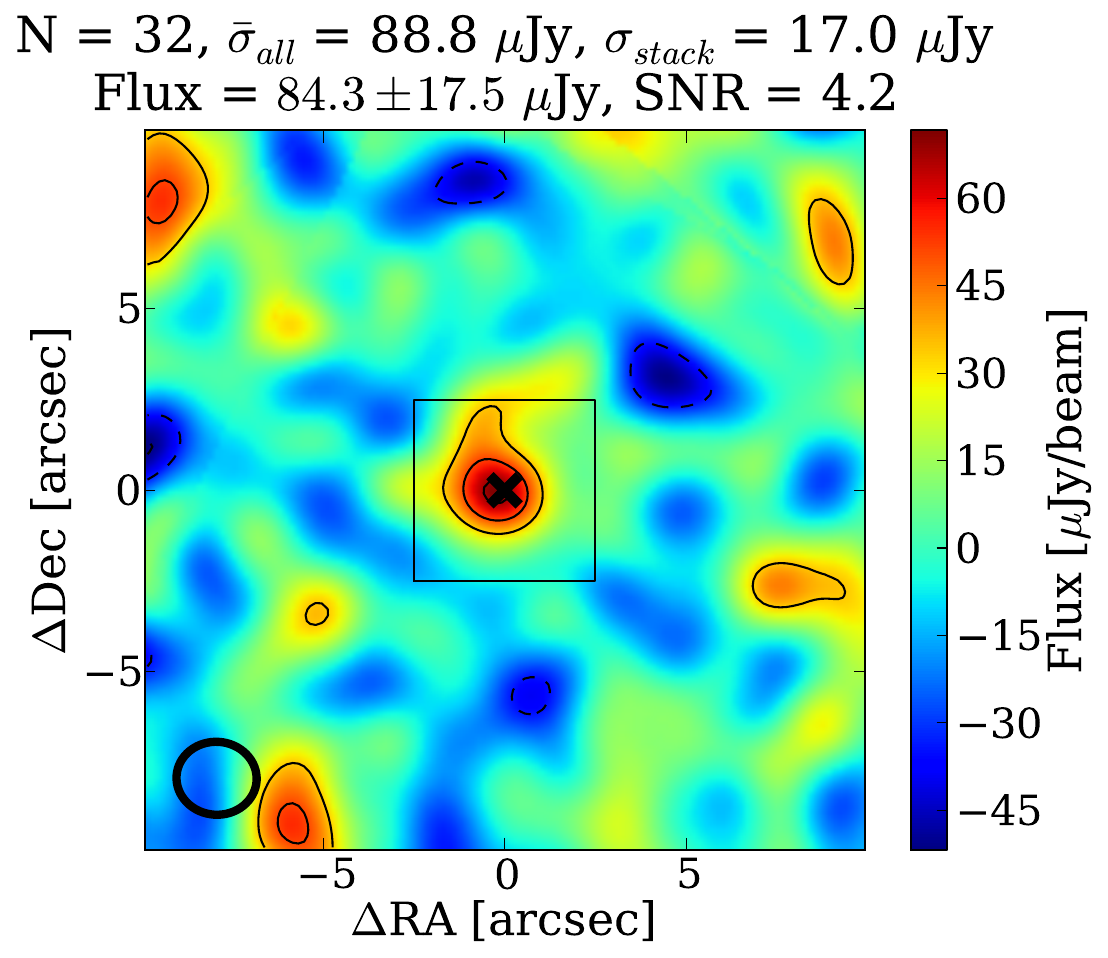}
\end{subfigure}\hfil
\begin{subfigure}{0.25\textwidth}
  \includegraphics[width=\linewidth]{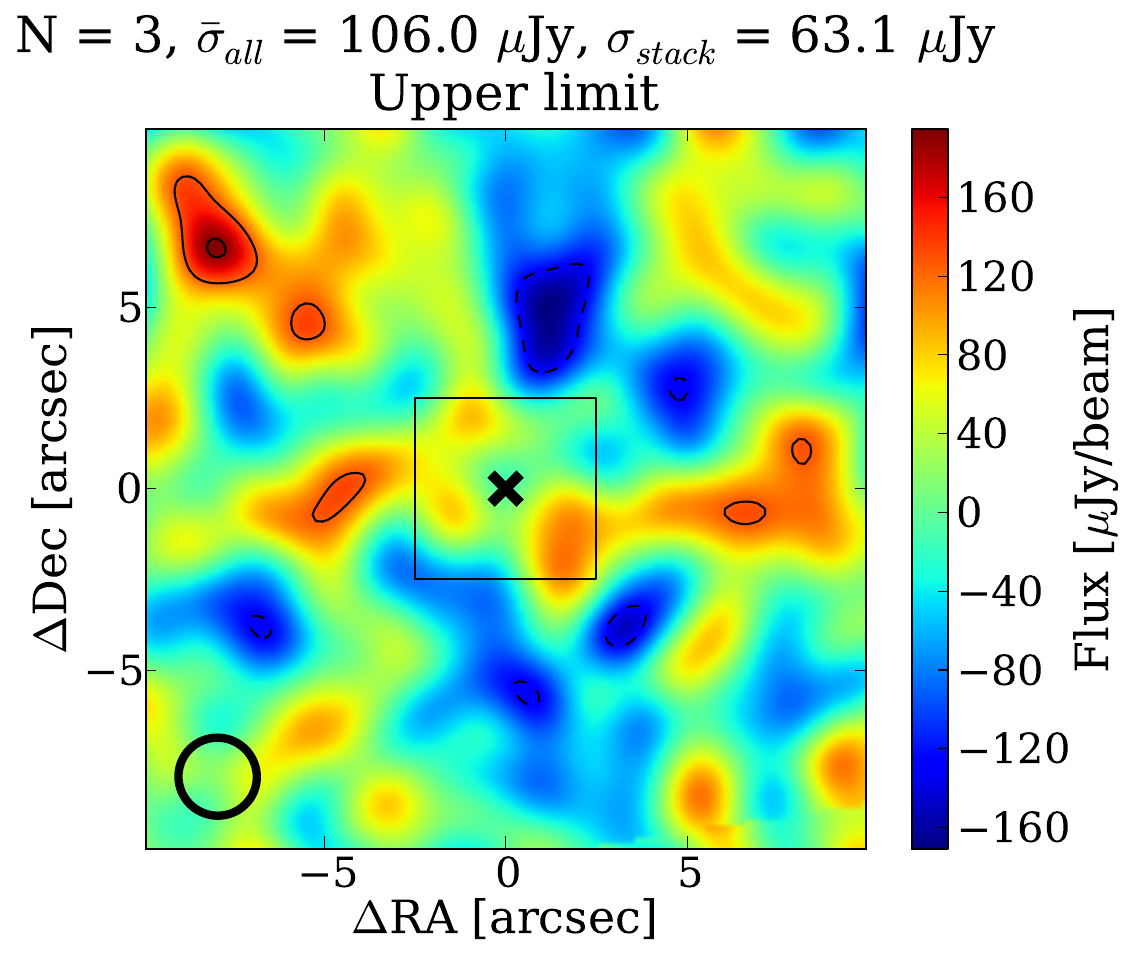}
\end{subfigure}
\medskip

\settoheight{\tempdim}{\includegraphics[width=0.25\textwidth]{f_z1m1.pdf}} 
\rotatebox{0}{\makebox[\tempdim]{\Large $z_2$}}\hfil
\begin{subfigure}{0.25\textwidth}
  \includegraphics[width=\linewidth]{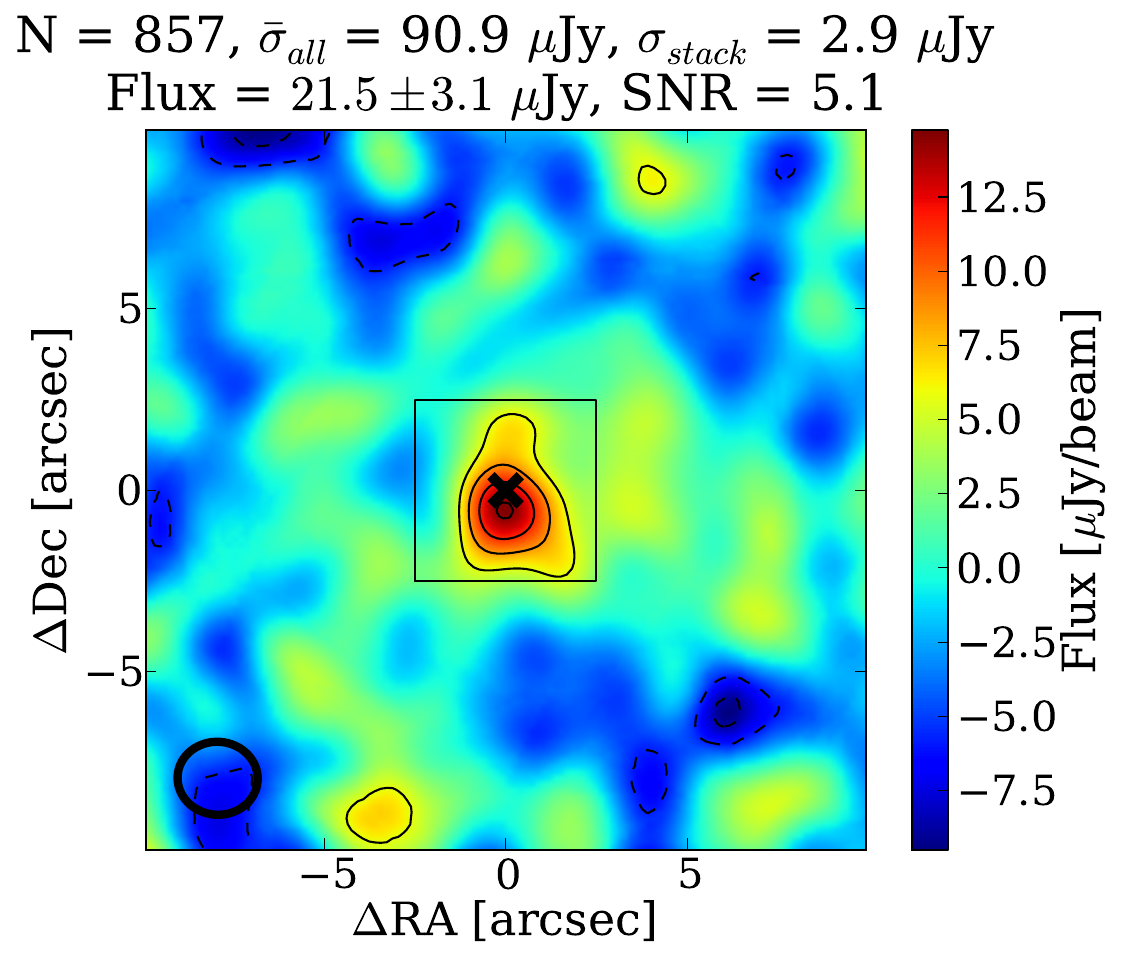}
\end{subfigure}\hfil 
\begin{subfigure}{0.25\textwidth}
  \includegraphics[width=\linewidth]{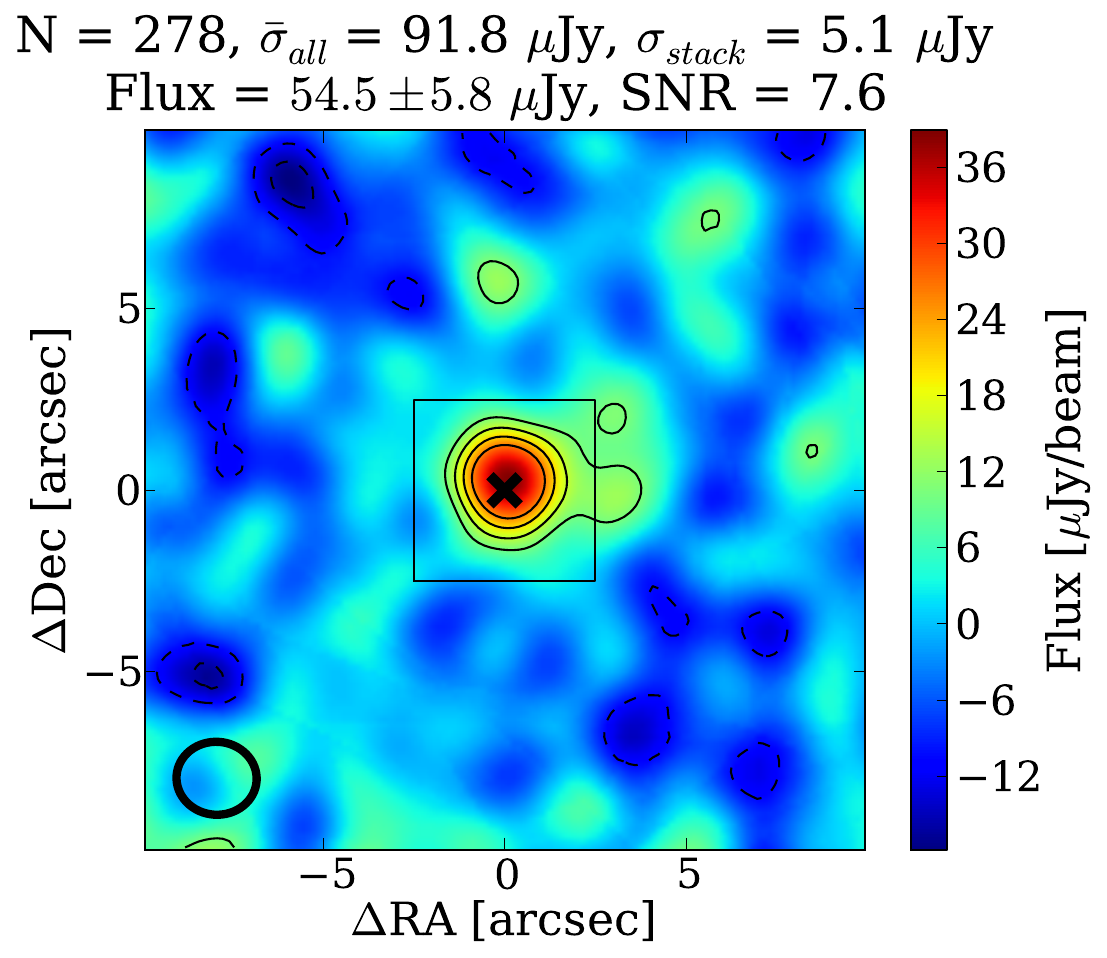}
\end{subfigure}\hfil  
\begin{subfigure}{0.25\textwidth}
  \includegraphics[width=\linewidth]{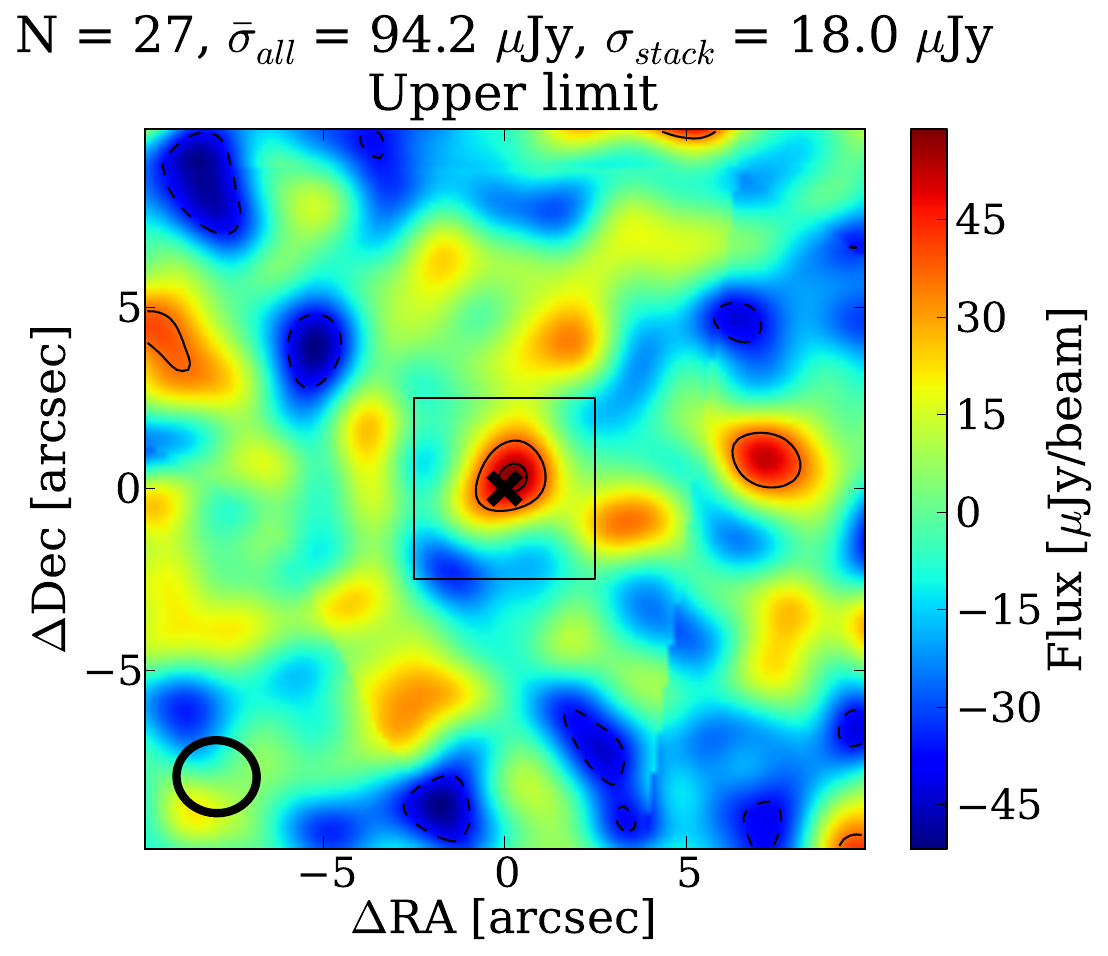}
\end{subfigure}
\medskip

\settoheight{\tempdim}{\includegraphics[width=0.25\textwidth]{f_z1m1.pdf}} 
\rotatebox{0}{\makebox[\tempdim]{\Large $z_3$}}\hfil
\begin{subfigure}{0.25\textwidth}
  \includegraphics[width=\linewidth]{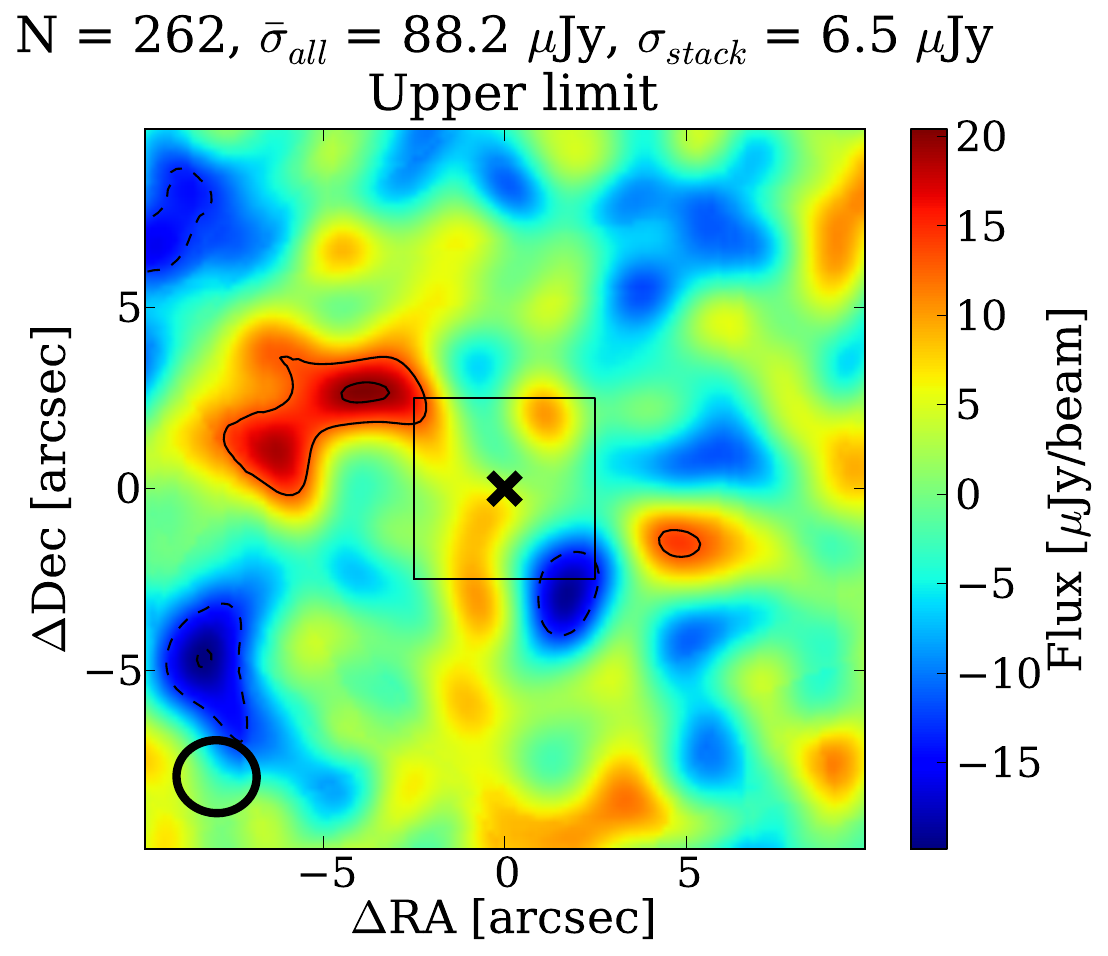}
\end{subfigure}\hfil 
\begin{subfigure}{0.25\textwidth}
  \includegraphics[width=\linewidth]{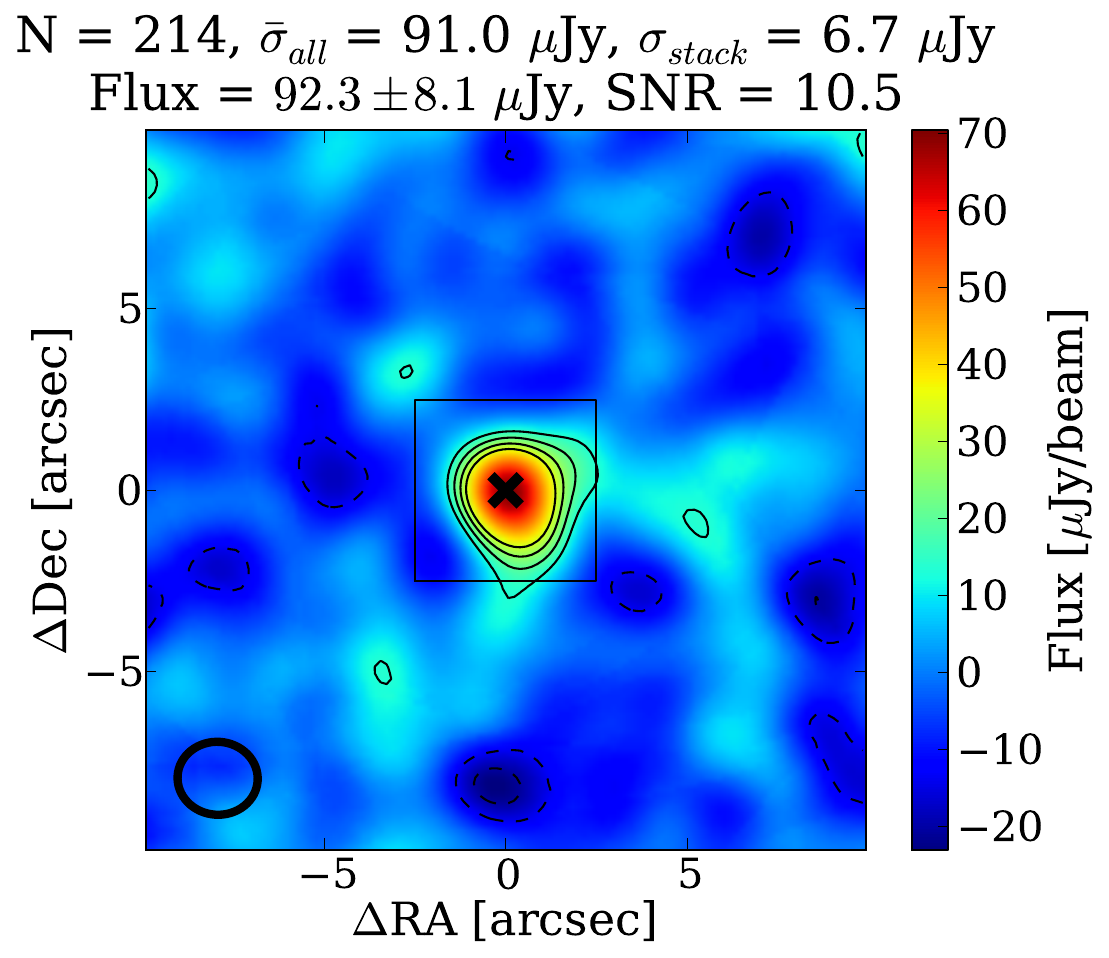}
\end{subfigure}\hfil  
\begin{subfigure}{0.25\textwidth}
  \includegraphics[width=\linewidth]{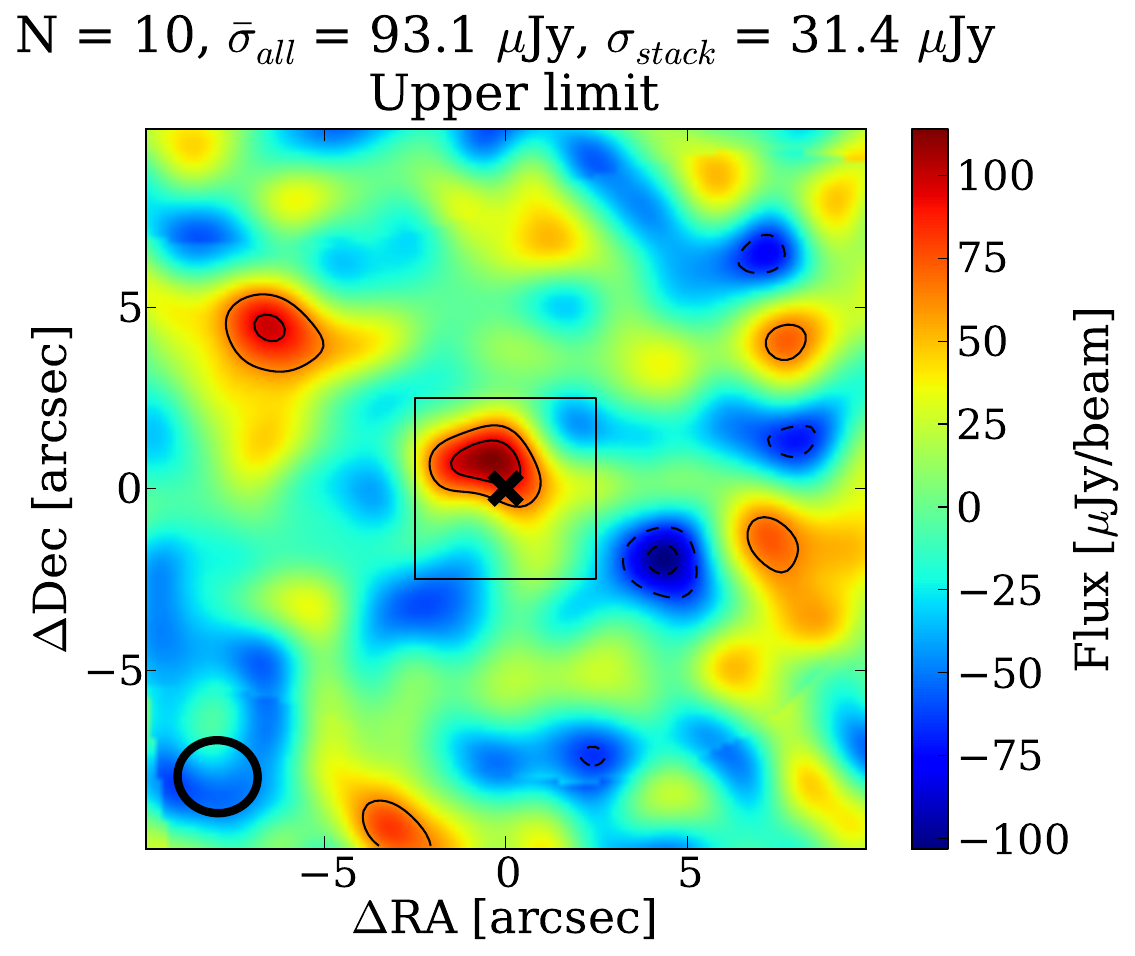}
\end{subfigure}
\caption{As in Fig. \ref{fig:cluster-bins}, but showing the stacked maps for field galaxies.
The rows (top to bottom) correspond to the redshift bins $0<z_1\leq0.4$, $0.4<z_2\leq1.0$ and $1.0<z_3\leq1.6$, respectively.}
\label{fig:stack-median-field}
\end{figure*}

\label{lastpage}
\end{document}